\documentclass{appolb}
\usepackage{epsfig}
\usepackage{graphicx}
\usepackage{dcolumn}   % needed for some tables
\usepackage{bm}        % for math
\usepackage{amsmath}
\usepackage{amssymb}   % for math
\usepackage{enumitem}
\usepackage{graphicx,mathrsfs}
\usepackage{nicefrac}
\usepackage{multirow}
\usepackage{color}

\newcommand{\rr}{\mbox{\boldmath $r$}}

\newcommand{\rb}{\mbox{\boldmath $b$}}
\newcommand{\rd}{\mbox{\boldmath $\Delta$}}

\newcommand{\pom}{\tt I\! P}
\def\xp{x_{{I\!\!P}}}
\def\lapproxeq{\lower .7ex\hbox{$\;\stackrel{\textstyle
<}{\sim}\;$}}
\def\gapproxeq{\lower .7ex\hbox{$\;\stackrel{\textstyle
>}{\sim}\;$}}
\def\funp{{I\!\!P}}
\def\xpom{x_{{I\!\!P}}}

\newcommand{\be}{\begin{equation}}
\newcommand{\ee}{\end{equation}}
\newcommand{\beqn}{\begin{eqnarray}}
\newcommand{\eeqn}{\end{eqnarray}}

\def\funp{{I\!\!P}}
\begin{document}
\title{Phenomenology of hard diffraction at high energies
\thanks{Presented at the Summer School and Workshop on High Energy Physics at
the LHC: New trends in HEP, October 21- November 6 2014,
Natal, Brazil}
}
\author{
Magno V. T. Machado
\address{High Energy Physics Phenomenology Group, GFPAE  IF-UFRGS \\
Caixa Postal 15051, CEP 91501-970, Porto Alegre, RS, Brazil}
}
\maketitle
\begin{abstract}
In this contribution we give a brief review on the application of perturbative QCD to the hard diffractive processes. Such reactions involving a hard scale can be  understood in terms of quarks and gluons degrees of freedom and have become an useful tool for investigating the low-$x$ structure of the proton and the behavior of QCD in the high-density regime. We start using the information from the $ep$ collisions at HERA concerned to the inclusive diffraction to introduce the concept of diffractive parton distributions. Their interpretation in  the resolved pomeron model is addressed and we discuss the limits of diffractive hard-scattering factorization for hadron-hadron collisions. Some examples of phenomenology for the diffractive production of $W/Z$, heavy $Q\bar{Q}$ and quarkonium in hadron-hadron reactions are presented.  We also discuss the exclusive diffractive processes in $ep$ interactions. They are in general driven by the gluon content of proton which is strongly subject to parton saturation effects in the very high energy limit.  These saturation effects are well described within the color dipole formalism. We present some examples of corresponding phenomenology as  the elastic vector meson production and the DVCS relying on the color dipole approach.
\end{abstract}


\section{Introduction}

Diffractive scattering envolves a large area of study in particle physics and gives rise to a wide range of
theoretical approaches.  Several aspects of diffraction in electron-proton collisions can be successfully
described in QCD if a hard scale (large photon vituality, heavy quark/quarkonia masses, large transverse momentum of particles) is present.  An important ingredient is the use of factorization theorems, which render parts of the dynamics
accessible to calculation in perturbation theory.  Namely, hard physics is associated with the  well established
parton picture and perturbative QCD. The remaining
non-perturbative quantities, as the diffractive PDFs can be extracted from measurements and
contain specific information about small-$x$ partons in the proton that
can only be obtained in diffractive processes. In first part of this contribution we will review the main features of diffractive deep inelastic scattering, where there are abundant and precise data which allow to explore the transition from hard to soft physics. On the other hand, for the  hard diffractive hadron-hadron collisions the situation is more
evolved since factorization is broken by rescattering between
spectator partons which are related with multiple
scattering effects. We will give some examples of phenomenology using the resolved pomeron model and simplified absorption corrections (the gap survival probability) for the diffractive production of heavy electroweak bosons, heavy quarks and quarkonia production in $p\bar{p}$ and $pp$ collisions of collider energies. We quote the review papers \cite{rp1,rp2,rp3,ADDIFF,GBWDIFFRA} and textbooks \cite{tb1,tb2,tb3} to the reader interested in a deeper analysis of the soft/hard diffraction phenomena in hadron-hadron and lepton-hadron collisions. Notice that our introductory text is strongly based on the celebrated papers quoted in Refs. \cite{rp1,rp2,rp3,ADDIFF,GBWDIFFRA}.

In the second part of this contribution, we discuss the diffractive exclusive processes in $ep$ collisions. We analyse the combination of data on inclusive and diffractive $ep$ scattering  and their connection to the test the onset of parton saturation at HERA. In particular, the diffractive vector meson production and deeply virtual Compton scattering (DVCS) have been extensively studied at HERA and provide a valuable  probe of the  QCD dynamics at high energies. In a general way, these processes are driven by the gluon content of target (proton or nuclei) which is strongly subject to parton saturation effects as well as considerable nuclear shadowing corrections when one considers scattering on nuclei. The cross sections for exclusive processes in DIS are proportional to the square of scattering amplitude, which turn it  strongly sensitive to the underlying QCD dynamics. They have been successfully described using color dipole approach and phenomenological model inspired in general aspects of parton saturation physics. We give some examples of the corresponding phenomenology using those approaches. We quote the review papers \cite{rp4,rp5,rp6,rp7,rp8} and textbook \cite{tb4} for a pedagogical treatment of these topics.

\section{Regge phenomenology for hadron interactions}

In hadron-hadron scattering an important fraction of the total cross
section is due to diffractive reactions. Examples of them are elastic
scattering (where both projectiles emerge intact in the final state), the single or double diffractive dissociation (where one or both
of hadrons are scattered into a low-mass state). A general feature of such diffractive processes  is the two groups
of final-state particles being well separated in phase space and in particular have a large gap in rapidity (LRG) between them.
Therefore, classical definition of diffraction in hadron-hadron or (virtual) photon-hadron scattering is the quasi
elastic scattering of one hadron combined with the dissociation of the second hadron or photon.

Diffractive hadron-hadron scattering can be described within Regge
theory \cite{Collins:1977jy}, which it was developed in the 1960s and  predates the
theory of the strong interactions, QCD. In this framework, the
exchange of particles in the $t$-channel is summed coherently to give
the exchange of so-called {\it regge trajectories}.  At sufficient high energies, diffraction is
characterized by the exchange of a specific trajectory, the
{\it Pomeron}, which has the quantum numbers of the vacuum.  Afterwards, it was found that
QCD perturbation theory in the high-energy limit can be organized
following the general concepts of Regge theory, referred to as BFKL formalism \cite{Kuraev:1976ge}.

In Regge theory the basic idea is that sequences of hadrons of mass $m_i$ and spin
$j_i$ lie on Regge trajectories $\alpha (t)$ such that $\alpha (m_i^2) = j_i$.   The corresponding Regge phenomenology is able to successfully describe all kinds of {\it soft} high energy hadronic scattering data:  differential, elastic and total cross section
measurements.  The high energy behaviour of a hadron scattering amplitude at
small angles ($t\rightarrow 0$) has the form
\begin{eqnarray}
\label{eq:a8}
A (s, t) \; \sim \; \sum_R \: \beta (t) \: s^{\alpha_R (t)}
\end{eqnarray}
Here,  $s$ is the square of the
centre-of-mass energy and $t$ is the square of the four-momentum transfer.  The observed
hadrons were found to lie on trajectories $\alpha_R (t)$ which are approximately linear in $t$.  The leading such
trajectories are the $\rho, a_2, \omega$ and $f$ trajectories which are all approximately
degenerate with \cite{DL}
\begin{eqnarray}
\label{eq:a10}
\alpha_R (t) \; \simeq \; 0.5 \: + \: 0.9 t.
\end{eqnarray}

From experimental point of view, the total cross sections are observed to increase slowly with energy at high
energies. Thus, one needs a higher lying trajectory as we can see using the optical theorem. This theorem expresses the total cross section for the process $AB \rightarrow X$ in terms of the imaginary part of the forward  elastic scattering amplitude ($AB \rightarrow AB$):
\begin{eqnarray}
\label{eq:b10}
\sigma (AB \rightarrow X) \; = \; \frac{1}{s} {\rm Im} A (s, 0) \; = \; \sum_R \:
\beta_R \: s^{\alpha_R (0) - 1}.
\end{eqnarray}

To account for the $s\rightarrow \infty$ dependence of the total cross sections a
Pomeron trajectory is invoked with intercept $\alpha_\funp (0)
\sim 1.08$. This Regge Pomeron is often called the {\it soft Pomeron}.  The total, elastic and differential hadronic cross section data are found to
be well described in the small-$t$ limit by taking a universal pole form for the Pomeron
\begin{eqnarray}
\label{eq:a9}
\alpha_\funp (t) \; \simeq \; 1.08 \: + \: 0.25 t,
\end{eqnarray}
plus the other sub-leading trajectories as in Eq. (\ref{eq:a10}) \cite{DL}. As a remark, the Pomeron
should be regarded as an effective trajectory, since the corresponding power behaviour on energy of the
total cross sections will ultimately violate the Froissart bound.

The effort in understanding diffraction in QCD has reached significant progress from studies of diffractive events \cite{HERAd}  at the
$ep$ collider HERA ($E_{e^{\pm}} \simeq 27.5$ GeV and $E_p\simeq 920$ GeV). The virtual photons/gauge bosons produced in these interactions can provide a hard scale where perturbative QCD methods can be applied. Several aspects of diffracion are well understood in QCD when a hard scale is present and then the dynamics can be formulated in the language of quarks and gluons. The possibility at HERA to scan a very large interval of photon virtualities allows to investigate what happens towards the non-perturbative region. This brings information on the soft diffractive processes as well.

In order to apply the Regge phenomenology to inclusive deep-inelastic scattering (DIS) and especially to its diffractive
component one makes use of the generalized optical theorem (Mueller's theorem \cite{MOT}).  The optical theorems
express the total cross sections in terms of the imaginary parts of the 2-body (or 3-body)
forward elastic
scattering amplitudes, or to be precise the discontinuities of the amplitudes across
the cuts along the $W^2$ (or $M^2$) axes. In the case of the inclusive DIS the  optical theorem gives
\begin{eqnarray}
\label{eq:a11}
F_2 \: \propto \sum_i \beta_i (W^2)^{\alpha_i (0) - 1} \quad \propto
\quad \sum_i \beta_i x^{1 - \alpha_i (0)}
\end{eqnarray}
for small $x$, see (\ref{eq:a3}).  In the naive parton model the valence and
sea quark contributions to $F_2$ are associated with meson and Pomeron exchange
respectively, and so using (\ref{eq:b3}) we have $xq_V \propto x^{1 - \alpha_R (0)} \propto x^{0.5},\,\, x q_S \propto x^{1 - \alpha_\funp (0)} \propto  x^{-0.08}$ for small-$x$ values.

For diffractive DIS, $\gamma^* p \rightarrow X p$, one applies Mueller's optical theorem \cite{MOT}.  In the limit of large $s/M^2$, the cross section is given by the discontinuity across the $M^2$ cut of the (three-body) $\gamma^* p\bar{p}$ elastic
amplitude, where a sum over the exchange Reggeons is implied.  The Regge prediction depends on whether $M^2$ is large or small.  For small $M^2$ the quark box gives the main contribution to photon-Pomeron scattering. Assuming $C = 1$ vector current coupling of the Pomeron
to quarks Ref.~\cite{DLdiff}, the resulting contribution to diffraction is found to be
\begin{eqnarray}
\label{DL-eq}
F_2^D \; \sim \; \beta (1 - \beta).
\label{eq:a13}
\end{eqnarray}

On the other hand, for large $M^2$ one has the double Regge limit ($s/M^2 \rightarrow
\infty$ and $M^2 \rightarrow \infty$) and the diffractive structure function is described by a
sum of triple Regge diagrams
\begin{eqnarray}
\label{eq:a14}
F_2^D \: \sim \: \sum_{i,j,k} \beta_{ijk} \left ( \frac{s}{M^2} \right
)^{\alpha_j (t) + \alpha_k (t)} \: (M^2)^{\alpha_i (0)}.
\end{eqnarray}

The leading behaviour, which is given by the triple Pomeron contribution, is
\begin{eqnarray}
\label{eq:a15}
F_2^D \; \sim \; (M^2)^{\alpha_\funp (0) - 2 \alpha_\funp (t)} \; \sim \; 1/M^2.
\end{eqnarray}

In next section, we address the extraction of diffractive structure function at HERA and its interpretation in the Regge phenomenology and the corresponding factorization formalism for the diffractive DIS processes.

\section{Difractive DIS and diffractive parton distributions}

Let us consider the inclusive DIS, $ep \rightarrow eX$, where $X$ represents all the fragments of the proton which has been broken up by the high
energy electron.  The basic subprocess $\gamma^* p \rightarrow X$, which can be expressed in terms of two functions $F_2$ and $F_L$ which characterize the structure
of the proton.  These proton structure functions depend on two invariant variables, the {\it virtuality} of the photon $Q^2 \equiv -q^2$ and the Bjorken $x$-variable
\begin{eqnarray}
\label{eq:a3}
x \; \equiv \; \frac{Q^2}{2 p.q} \; = \; \frac{Q^2}{Q^2 + W^2},
\end{eqnarray}
where $p$ and $q$ are the four-momenta of the proton and virtual photon, respectively. The quantity $W$ is the total $\gamma^* p$ centre-of-mass energy. In the parton model, $x$ is the fraction of the proton's momentum carried by the quark struck by the virtual photon.  In this simple quark model $F_L = 0$ and
\begin{eqnarray}
\label{eq:b3}
F_2 \; = \; F_T \; = \; \sum_q \: e_q^2 \: xq (x)
\end{eqnarray}
is independent of $Q^2$.  The sum is over the flavours of quarks, with electric charge $e_q$
(in units of $e$) and distributions $q (x)$.  $F_{T,L}$ are the proton structure functions for
DIS by transversely, longitudinally polarised photons. In the parton-QCD model (including the QCD radiation from the valence quarks and gluon radiation) the parton distributions $q (x)=q(x,Q^2)$ acquire dependence on the hard scale associated to the process. They are now evoluted by evolution equations on the virtuality $Q^2$ (the DGLAP equations \cite{DGLAP}).

The general form of the DIS cross section, up to target mass corrections, is
\be
\label{eq:a4}
\frac{d^2 \sigma (ep \rightarrow eX)}{dxdQ^2} \; = \; \frac{2 \pi \alpha^2}{x Q^4} \: \left \{
[1 + (1 - y)^2] \: F_2 (x, Q^2) \: - \: y^2 F_L (x, Q^2) \right \}
\ee
where $\alpha$ is the electromagnetic coupling.  The third variable $y$ is needed to fully
characterize the DIS process, $ep \rightarrow eX$, namely $y = Q^2/xs$ where
$\sqrt{s}$ is the total centre-of-mass energy of the electron-proton collision.

Now, in a typical diffractive event at HERA the collision of the virtual photon with the proton produces a
  hadronic final state $X$ with the photon quantum numbers and
  invariant mass $M_X$. A large gap in rapidity is
  present between $X$ and the final-state
  proton, which emerges with its momentum barely changed. Diffractive DIS thus combines features of hard and soft
scattering.  The kinematics of $\gamma^* p \to Xp$ can be described by the
invariants $Q^2 =-q^2$ and $t =(p-p')^2$, and by the scaling variables
$x_{\pom}$ and $\beta$ given by
\begin{equation}
  \label{xpom-def}
x_{\pom} = \frac{(p-p')\cdot q}{P\cdot q} =
\frac{Q^2+M_X^2-t}{W^2+Q^2-M_p^2} ,
\qquad
\beta = \frac{Q^2}{2(p-p')\cdot q} = \frac{Q^2}{Q^2+M_X^2-t} ,
\end{equation}
where $W^2 =(p+q)^2$. The variable $x_{\pom}$ is the
fractional momentum loss of the incident proton.  The quantity
$\beta$ has the form of a Bjorken variable defined with respect to the
momentum $p-p'$ lost by the initial proton instead of the initial
proton momentum $p$. The usual Bjorken variable
$x = Q^2 /(2 p\cdot q)$ is related to $\beta$ and $x_{\pom}$ as
$\beta x_{\pom} = x$.

The cross section for $ep \to eXp$ in the one-photon exchange
approximation can be written in terms of diffractive structure
functions $F_2^{D(4)}$ and $F_L^{D(4)}$ as
\begin{equation}
\frac{d\sigma^{4}(ep \rightarrow eXp)}{
d\beta\, dQ^2\,dx_{\pom}\,dt} =
\frac{4\pi\alpha^2}{\beta Q^4}
\biggl[ \Big(1-y+\frac{y^2}{2}\Big) F_2^{D(4)}(\beta,Q^2,x_{\pom},t)
  - \frac{y^2}{2} F_L^{D(4)}(\beta,Q^2,x_{\pom},t)
\biggr] ,
\label{sigma-2}
\end{equation}
in analogy with the way $d\sigma (ep \rightarrow eX)/(dx\, dQ^2)$
is related to the structure functions $F_2$ and $F_L$ for inclusive DIS,
$ep\to eX$. Here $y=(p\cdot q)/(p\cdot k)$ is the fraction of
energy lost by the incident lepton in the proton rest frame.  The
structure function $F_L^{D(4)}$ corresponds to longitudinal
polarization of the virtual photon; its contribution to the cross
section is small in a wide range of the experimentally accessible
kinematic region (in particular at low $y$).  The structure function
$F_2^{D(3)}$ is obtained from $F_2^{D(4)}$ by integrating over~$t$:
\begin{equation}
F_2^{D(3)}(\beta, Q^2,x_{\pom})=\int dt\,
F_2^{D(4)}(\beta,Q^2,x_{\pom},t)  .
\label{f2d3}
\end{equation}

In a parton model picture, inclusive diffraction $\gamma^* p \to Xp$
proceeds by the virtual photon scattering on a quark, in analogy to
inclusive scattering.  In this
picture, $\beta$ is the momentum fraction of the struck quark with
respect to the exchanged momentum $p-p'$.  The diffractive
structure function describes the proton structure in these specific
processes
with a fast proton in the final state. $F_2^D$ may also be viewed as
describing the structure of whatever is exchanged in the $t$-channel
in diffraction. In the Regge language this is the exchange of a Pomeron if multiple regge exchange can
be neglected.  However, the Pomeron in QCD (for instance, the two-gluon exhange model) cannot be interpreted as a particle on which the
virtual photon scatters. Using the QCD factorization theorem for inclusive diffraction, $\gamma^* p\to Xp$, the diffractive
structure function, in the limit of large $Q^2$ at fixed $\beta$,
$\xpom$ and $t$, can be written as \cite{Trentadue:1993ka,Berera:1995fj,Collins:1997sr}
\begin{equation}
\label{eq:dpd}
F_{2}^{D(4)}(x, Q^2, x_{\pom}, t)\,=\,
\sum_a\,\int_0^{x_{\pom}} d\xi\;
{\cal{F}}^D_{a/p}(\xi, \mu^2,\xp, t)\;
{\cal{C}}_{a}(x/\xi,Q^2/\mu^2)\,,
\end{equation}
with $a=q,g$ denoting a quark or gluon distribution in the proton,
respectively. In the infinite momentum  frame  the diffractive
parton distributions describe the probability to find a parton  with the
fraction $\xi$ of the proton momentum,  provided the proton stays intact and
loses only a small fraction $\xp$ of its original momentum. ${\cal{C}}_{a}$
are the  coefficient functions   describing hard scattering
of the virtual photon on a parton $a$.
They are identical to the   coefficient functions known from inclusive DIS,
\begin{equation}
\label{eq:hcs}
{\cal{C}}_{a}(x/\xi,Q^2/\mu^2)\,=\,e_a^2\,\delta(1-x/\xi)\,+\,
{\cal{O}}(\alpha_{s})\,.
\end{equation}

Formula in Eq. (\ref{eq:dpd}) is the analogue of the inclusive leading twist
description for inclusive DIS. The scale $\mu^2$ is the factorization/renormalization scale and we notice that since
the l.h.s of  Eq.~(\ref{eq:dpd}) does not depend on this scale ($d F_2^{D(4)}/d\mu^2=0$), one finds the
renormalization group equations for the diffractive
parton distribution
\begin{equation}
\label{eq:ap}
\mu^2\,\frac{d}{d\mu^2}\;
{\cal{F}}^D_{a/p}(\xi, \mu^2,\xp, t)
\,=\,
\sum_b \int_{\xi}^{\xp} \frac{dz}{z}\,P_{a/b}(\xi/z,\alpha_s(\mu^2))\;
{\cal{F}}^D_{b/p}(z, \mu^2,\xp, t)\,,
\end{equation}
where $P_{a/b}$ are the standard Altarelli-Parisi splitting functions
in leading (LO) or next-to-leading (NLO) logarithmic approximation. Since the
scale $\mu$ is arbitrary, we can choose $\mu=Q\gg\Lambda_{QCD}$. If we refer the longitudinal momenta of the partons to
$\xp p$ instead of the proton total momentum $p$, the structure functions
and parton distributions become functions of $\beta=x/\xp$ or
${\beta}^\prime=\xi/\xp$. Using this notation, one rewrites Eqs. (\ref{eq:dpd}) and
(\ref{eq:ap}) in the following way:
\begin{equation}
\label{eq:dpd1}
F_{2}^{D(4)}(\beta, Q^2, \xp, t)\,=\,
\sum_a\,\int_0^{1} d{\beta}^\prime\;
\xp{\cal{F}}^D_{a/p}({\beta}^\prime, \mu^2,\xp, t)\;
{\cal{C}}_a(\beta/{\beta}^\prime,Q^2/\mu^2)\,
\end{equation}
and
\begin{equation}
\label{eq:ap1}
\mu^2\,\frac{d}{d\mu^2}\;
{\cal{F}}^D_{a/p}({\beta}, \mu^2,\xp, t)
\,=\,
\sum_b \int_{{\beta}}^1
\frac{dz}{z}\,P_{a/b}({\beta}/z,\alpha_s(\mu^2))\;
{\cal{F}}^D_{b/p}(z,\mu^2,\xp, t)\,.
\end{equation}

Thus, we obtain a description similar to inclusive DIS but modified
by the additional variables $\xp$ and $t$. Moreover, the Bjorken variable $x$
is replaced by its diffractive analogue $\beta$. Notice that $\xp$ and $t$ play the role of parameters
of the evolution equations and  does not affect
the evolution. According to the factorization theorem
the evolution equations (\ref{eq:ap1}) are applicable to all orders
in perturbation theory. In LO approximation for the coefficient functions
(\ref{eq:hcs}), one finds for the diffractive structure function (summing over the quark flavours)
\begin{equation}
\label{eq:llsf}
F_{2}^{D(4)}(\beta, Q^2, \xp, t)\,=\,\sum_{a=q,\bar{q}}\,e_a^2\;
\beta\;
\xp{\cal{F}}^D_{a/p}({\beta}, Q^2,\xp, t)\,,
\end{equation}
where the sum over the quark flavours is performed.

The collinear factorization formula (\ref{eq:dpd1})
holds to all orders in $\alpha_s$ for diffractive DIS
\cite{Collins:1997sr}. However, this is no longer true in
hadron--hadron hard diffractive scattering \cite{rp2,CFS93}, where collinear
factorization  fails due  to final state soft interactions. Thus, unlike
inclusive scattering,  the diffractive parton distributions are no universal
quantities. The can safely be used, however, to describe hard
diffractive processes involving leptons.

Using a Regge language, in the resolved pomeron model (Ingelman-Schlein model \cite{IS}) diffraction is described with the
help of the concept of pomeron exchange. It is assumed
that the pomeron has a hard structure and in DIS diffraction this structure would be resolved by the virtual photon. Thus, the resolved pomeron model
is based on the assumption of {\it Regge factorization}. In this picture the diffractive structure function takes a factorized form $F_2^{D(4)} = f_{\pom}\, F_2^{\pom}$, where $f_{\pom}$ is the {\it Pomeron flux} describing   the emission of the Pomeron from the proton and its subsequent propagation, and where $F_2^{\pom}$ is the {\it pomeron structure function}.  Phenomenologically, such a factorizing ansatz works not too badly and is often used. In the context of the
diffractive parton distributions it means that the following factorization
holds \cite{BERSOP94,BERSOP96}
\begin{equation}
\label{eq:is}
\xp{\cal{F}}^D_{a/p}({\beta}, Q^2,\xp, t)\,=\,
f(\xp,t)\ f_{a/\funp}({\beta},Q^2)\,,
\end{equation}
where the pomeron flux $f(\xp,t)$ is given by
\begin{equation}
f(\xp,t)\,=\,
\frac{F^2(t)}{8\pi^2}\;\xp^{1-2\alpha_\funp(t)}\,.
\end{equation}
Thus, the variables $(\xp,t)$,
related to the loosely scattered proton, are factorized from
the variables characterizing the diffractive system $(\beta,Q^2)$.
$F(t)$ is the Dirac electromagnetic form factor \cite{DLdiff},
$\alpha_\funp(t)=1.1+0.25~\mbox{\rm GeV}^{-2}\cdot t$ is the soft pomeron
trajectory \cite{DL} and
the normalization  of  $f(\xp,t)$ follows the
convention of \cite{DLdiff}. The function $f_{a/\funp}(\beta,Q^2)$
in Eq.~(\ref{eq:is}) describes the
hard structure in DIS diffraction, and  is interpreted
as the pomeron parton distribution.
Now, the diffractive
structure function (\ref{eq:llsf}) becomes
\begin{equation}
\label{eq:isfd}
F_{2}^{D(4)}(\beta, Q^2, \xp, t)\,=\,f(\xp,t)
\;
\sum_{a=q,\overline{q}}\,e_a^2\,\beta\,f_{a/\funp}(\beta,Q^2)\,,
\end{equation}
where the summation over quarks and antiquarks is performed.
The  $Q^2$-evolution of $f_{a/\funp}(\beta,Q^2)$ is given by the
DGLAP equations (\ref{eq:ap1}). The $t$-dependence in the pomeron
parton distributions is neglected. The pomeron parton distributions are determined  as the parton
distributions of  real hadrons. Some functional form with several
parameters  is assumed at an initial scale
 and then the  parameters are found from a fit to data
\cite{ROY00,H197,ZEUS99,H1FPS} using the DGLAP evolution equations.

Despite the success for describing diffractive DIS and related processes in $ep$ collisions the diffractive hard-scattering factorization does not apply to hadron-hadron collisions~\cite{Berera:1995fj,Collins:1997sr}. The discrepancy is quite large as the fraction of diffractive dijet events at CDF is a factor 3 to 10
smaller than would be expected on the basis of the HERA data \cite{cdf}. The same
type of discrepancy is consistently observed in all hard diffractive
processes in $p\bar{p}$ events, see e.g.~\cite{Alvero:1998ta}.  In
general, while at HERA hard diffraction contributes a fraction of order
10\% to the total cross section, it contributes only about 1\% at the
Tevatron. Attempts to establish corresponding factorization theorems fail
 because of interactions between spectator partons of the colliding
   hadrons.  The contribution of these interactions to the cross section
   does not decrease with the hard scale.  Since they are not associated
   with the hard-scattering subprocess, we no
   longer have factorization into a parton-level cross section and the
   parton densities of one of the colliding hadrons. These
interactions are generally soft, and we have at present to rely on
phenomenological models to quantify their effects \cite{GLMrev}. The yield of diffractive events in hadron-hadron collisions is lowered
precisely because of these soft interactions between spectator partons.
They can produce additional final-state particles which fill the would-be
rapidity gap. This is the season for the often terminology {\it rapidity gap survival}.  When
such additional particles are produced, a very fast proton can no longer
appear in the final state because of energy conservation.  Diffractive
factorization breaking is thus intimately related to multiple scattering
in hadron-hadron collisions.

In next section, we give some examples of phenomenology of hard diffraction in hadron-hadron collisions using the resolved pomeron model supplemented by {\it rapidity gap survival} corrections for some representative processes as heavy electroweak boson, heavy quarks and quarkonia production in Tevatron and LHC energies.

%%%%%%%%%%%%%%%%%%%%%%%%%%%%%%%%%%%%%%%%%%%%%%%%%%%%%%%%%%%%%

\section{Some examples of phenomenology in proton-proton collisions}
\label{samplespphd}

One of the main baseline process in hard diffraction is the production of heavy gauge bosons. In what follows we summarize the results obtained in \cite{ggm_prd}, where the diffractive $W$ and $Z$ production are computed for the Tevatron energy and estimates are provided for the CERN LHC experiment. For the hard diffractive processes we will consider the resolved-pomeron picture \cite{IS} where
the Pomeron structure is probed as discussed in previous section. The generic cross section for a process in which partons of two hadrons, $A$ and $B$, interact to produce a massive electroweak boson, $ A + B \rightarrow W^{\pm} + X$, reads as
\begin{eqnarray}
\frac{d \sigma}{dx_a\,dx_b} = \sum_{a,b} f_{a/A}(x_a,
\mu^2)\, f_{b/B}(x_b, \mu^2)\, \frac{d\hat{\sigma}(ab\rightarrow W(Z)\,X)}{d\hat{t}}\,,
\label{gen}
\end{eqnarray}
where $x_i f_{i/h}(x_i, \mu^2)$ is the distribution function of a parton of flavour $i=a,b$ in the hadron $h=A,B$.  The quantity $d\hat{\sigma}/d\hat{t}$ gives the elementary hard cross section of the corresponding subprocess and $\mu^2=M_{W}^2$ is the hard scale in the QCD evolution. In the expression for diffractive processes, one assumes that one of the hadrons, say hadron $A$, emits a Pomeron whose partons interact with partons of the hadron $B$. Thus the parton distribution  $x_a f_{a/A}(x_a, \mu^2)$ in Eq.~(\ref{gen}) is replaced by the convolution between a
distribution of partons in the Pomeron, $\beta f_{a/{\pom}}(\beta, \mu^2)$, and the ``emission rate" of Pomerons by the hadron, $f_{{\pom}/h}(x_{{\pom}},t)$. The last quantity, $f_{{\pom}/h}(x_{{\pom}},t)$, is the Pomeron flux factor and its explicit formulation is described in terms of Regge theory. Therefore, we can rewrite the parton distribution as
\begin{eqnarray}
\label{convoP}
x_a f_{a/A}(x_a, \,\mu^2)\ =\ \int dx_{{\pom}} \
\bar{f}(x_{{\pom}})\, {\frac{x_a}{x_{{\pom}}}}\, f_{a/{\pom}}
({\frac{x_a}{x_{{\pom}}}}, \mu^2).
\end{eqnarray}
where we have defined the quantity $\bar{f} (x_{{\pom}}) \equiv \int_{-\infty}^0 dt\
f_{{\pom/A}}(x_{{\pom}},t)$.

Concerning the $W^{\pm}$ diffractive production, one considers the reaction
$p + {\bar p}(p) \rightarrow p + \ W (\rightarrow e\ \nu ) + \ X$, assuming that a Pomeron emitted by a proton in
the positive $z$ direction interacts with a $\bar p$ (or a $p$) producing $W^{\pm}$
that subsequently decays into $e^{\pm}\ \nu$. By using the same concept of the convoluted structure function, the
diffractive cross section for the inclusive lepton production becomes
\begin{eqnarray}
\frac{d\sigma^{\mathrm{SD}}_{\mathrm{lepton}}}{d\eta_e} & = &\sum_{a,b}
\int \frac{dx_{\pom}}{x_{\pom}}\, \bar{f}(x_{\pom})
\int dE_T \ f_{a/{\pom}}(x_a, \,\mu^2)\,f_{b/\bar{p}(p)}(x_b, \,\mu^2) \nonumber \\
& & \left[\frac{ V_{ab}^2\ G_F^2}{6\ s\ \Gamma_W}\right]\ \frac{\hat{t}^2}
{\sqrt{A^2-1}}
\label{dsw}
\end{eqnarray}
where
\begin{eqnarray}
x_a = \frac{M_W\ e^{\eta_e}}{(\sqrt{s}\ x_{{\tt I\! P}})}\ \left[A \pm
\sqrt{(A^2-1)}\right]
\label{xaw},\,\,\,x_b = \frac{M_W\ e^{-\eta_e}}{\sqrt{s}}\ \left[A \mp \sqrt{(A^2-1)}\right],
\end{eqnarray}
with $A={M_W}/{2 E_T}$, $E_T$ being the lepton transverse energy, $G_F$ is the Fermi constant and the hard scale $\mu^2=M_W^2$. The quantity  $V_{ab}$ is the Cabibbo-Kobayashi-Maskawa matrix element and $\hat{t}=-E_T\ M_W\ \left[A+\sqrt{(A^2-1)}\right]$. The upper signs in Eqs.~(\ref{xaw}) refer to $W^+$ production (that is, $e^+$ detection). The corresponding cross section for $W^-$ is obtained by using the lower signs and ${\hat t}
\leftrightarrow {\hat u}$.
The detection of this reaction is triggered by the leptons ($e^{+}$ for $W^{+}$ and $e^{-}$ for $W^{-}$) that appears boosted towards negative rapidity $\eta$ in coincidence with a rapidity gap in the right hemisphere.

Since the same concept, the cross section for the diffractive hadroproduction oh the boson $Z$ is given by
\begin{eqnarray}
\sigma_{Z} & = & \sum_{a,b}\int\frac{dx_{\pom}}{x_{\pom}}\int\frac{dx_{b}}{x_{b}}\int\frac{dx_{a}}{x_{a}}\bar{f}(x_{\pom})f_{a/\pom}(x_{a},\mu^{2})f_{b/\bar{p}(p)}(x_{b},\mu^{2}) \\ \nonumber & \times & \left [\frac{2\pi C^{Z}_{a,b}G_{F}M^{2}_{Z}}{3\sqrt{2}s}\right ]\frac{d\hat{\sigma}(ab\rightarrow ZX)}{d\hat{t}}\, ,
\label{equacaoZ}
\end{eqnarray}
where
\begin{eqnarray}
C^{Z}_{q\bar{q}}=\frac{1}{2}-2|e_{q}||sin^{2}\theta_{W}+4|e_{q}|^{2}sin^{4}\theta_{W}
\end{eqnarray}
with $\theta_{W}$ being the Weinberg angle. The definitions for $x_{a,b}$ are similar as for the $W$ case and now $\mu^{2}=M^{2}_{Z}$. $\theta_{C}=0.2269$ is the Cabibbo angle and the other values of the electroweak parameters are taken from the Particle Data Group \cite{pdg}.

As we discussed in previous section, the factorization does not
necessarily hold for diffractive production processes. The suppression
of the single-Pomeron Born cross section due to the multi-Pomeron  
contributions depends, in general, on the particular hard process.  We
will consider this suppression through a gap survival probability
factor, $S_{\mathrm{gap}}^2$, using two theoretical estimates for this factor: (a) model of \cite{KKMR} (labeled KMR), which considers a two-channel eikonal model. The survival probability is computed for single, central and double diffractive processes at several energies. We will consider the results for single diffractive processes, where $S_{\mathrm{gap}}^2 (KMR)=0.15$ for $\sqrt{s}=1.8$ TeV (Tevatron) and $S_{\mathrm{gap}}^2 (KMR)=0.09$ for $\sqrt{s}=14$ TeV (LHC). (b) The second theoretical estimate is from \cite{GLM} (labeled GLM), which considers a single channel eikonal approach, where $S_{\mathrm{gap}}^2(GLM)=0.126$ for $\sqrt{s}=1.8$ TeV (Tevatron) and $S_{\mathrm{gap}}^2(GLM)=0.081$ for $\sqrt{s}=14$ TeV (LHC).
\begin{table}
\begin{tabular}{cccc} 
\hline
$\sqrt{s}$ & Rapidity & Data ($\%$) & Estimate ($\%$)\\
\hline
1.8 TeV &    $|\eta_{e}|<1.1$  & $1.15\pm 0.55$ \cite{CDF}  & $0.715\pm 0.045$\\
1.8 TeV &    $|\eta_{e}|<1.1$  & $1.08\pm 0.25$ \cite{D0} & $0.715\pm 0.045$\\
1.8 TeV &    $1.5<|\eta_{e}|<2.5$  & $0.64\pm 0.24$  \cite{D0} & $1.700\pm 0.875$\\
1.8 TeV &     Total $W\rightarrow e\nu $ & $0.89\pm 0.25$ \cite{D0} & $0.735\pm 0.055$ \\
14 TeV &    $|\eta_{e}|<2$  & ---  & $0.311\pm 0.016$\\
\hline
\end{tabular}
\caption{Data versus model predictions for diffractive $W^{\pm}$ hadroproduction (cuts $E_{T_{\mathrm{min}}}=20$ GeV and $x_{\pom}<0.1$). }
\end{table}

Let us present some results for hard diffractive production of
W and Z based on the present discussion. They are compared
with experimental data from \cite{CDF,D0} in Table I, where estimates for the LHC are also presented. In the numerical calculations, we have used the new H1 parameterizations for the diffractive pdf's \cite{H1FPS}  As the larger uncertainty comes from the gap survival factor, the error in the predictions correspond to the theoretical band for $S_{\mathrm{gap}}^2$. In the theoretical expressions of previous section one computes only the interaction of pomerons (emmitted by protons) with antiprotons (protons in LHC case), that means events with rapidity gaps on the side from which antiprotons come from.  Disregarding the gap factor, the diffractive production rate is approximately 7 \% (using the cut $|\eta|<1$) being very large compared to the Tevatron data. When considering the gap survival probability correction, the values are in better agreement with data. When considering central $W$ boson fraction, $-1.1<\eta_e<1.1$ (cuts of CDF and D0 \cite{CDF,D0}), we obtain a diffractive rate of 0.67 \% using the KMR estimate for $S_{\mathrm{gap}}^2$, whereas it reaches 0.76\% for the GLM estimate. The average rate considering the theoretical band for the gap factor is then $R_W= 0.715\pm 0.045$ \%.  Considering the forward $W$ fraction, $1.5<|\eta_e|<2.5$ (D0 cut), one obtains $R_W=0.83$ \% for KMR and $R_W=2.58$ \% for GLM, with an averaged value of $R_W= 1.7\pm 0.875$ \%. In this case, our estimate is larger than the central experimental value $R_W^{\mathrm{D0}}=0.64$ \%. For the total $W\rightarrow e\nu$ we have $R_W=0.68$ \% for KMR and $R_W=0.79$ \% for GLM  and the mean value $R_W= 0.735 \pm 0.055$ \%, which is in agreement with data and consistent with a large forward contribution. Finally, we estimate the diffractive ratio for LHC energy, $\sqrt{s}=14$ TeV. In this case we extrapolate the pdf's in proton and diffractive pdf's in Pomeron to that kinematical region. This procedure introduces somewhat additional uncertainties in the theoretical predictions. We take the conservative cuts $|\eta_e|<2$, $E_{T_{\mathrm{min}}}=20$ GeV for the detected lepton and $x_{\pom}<0.1$. We find $R_W=0.327$ \% for KMR gap survival probability factor and $R_W=0.295$ \% for GLM, with a mean value of $R_W^{\mathrm{LHC}}=0.0311\pm 0.016 $ \%. The CMS Collaboration already has a signal for single diffractive boson production \cite{CMSWdiff} at 7 TeV, where the diffractive ratio was determined to be $0.73\pm 0.34$ \cite{CMSWdiff}.

We now refer to recent works on this topic. For instance, in Ref. \cite{GBA} the analysis of diffractive electroweak vector boson production was done and the author show that the single diffractive $W$ production asymmetry in rapidity is a good observable at the LHC to test the concept of the flavour symmetric pomeron parton distributions. Along these studies, in Ref. \cite{grss} has been shown taht double diffractive electroweak boson production is an ideal probe of QCD based mechanisms of diffraction. Namely, assuming the resolved pomeron model with flavour symmetric pdfs, the $W$ production asymmetry in rapidity equals zero at LHC. On the other hand, in the soft color interaction (SCI) model \cite{IPRW} that asymmetry is non-zero and it is similar to the asymmetry in the inclusive case. A discrepancy also occurs for the ratio $W/Z$, which is independent of rapidity in the resolved pomeron model and rapidity-dependent in SCI models. Finally, the diffractive production has been addressed also within the color dipole approach \cite{PKP}, where the introduction of higher twist contributions and breakdown of diffractive factorization  are naturally embeded.

The next example refers to the heavy quark production in single and double diffractive dissociation in hadron colliders. In what follows we summarize the results found in Refs. \cite{QQgmm1,QQgmm2,QQmvtm}. Let us present the main formulas for the inclusive diffractive cross sections for the production of heavy quarks in proton-proton collisions at high energies. In the inclusive case, the process is described for partons of two protons, interacting to produce a heavy quark pair, $p+p\rightarrow Q\bar{Q}+X$, with center of mass energy $\sqrt{s}$. At LHC energies, the gluon fusion channel dominates over the $q\bar{q}$ annihilation process and $qg$ scattering. The NLO cross section is obtained by convoluting the partonic cross section with the parton distribution function (PDF), $g(x,\mu_F)$, in the proton, where $\mu_F$ is the factorization scale. At any order, the partonic cross section may be expressed in terms of dimensionless scaling functions $f^{k,l}_{ij}$ that depend only on the variable $\rho$ \cite{13magno},
\begin{eqnarray}
\hat{\sigma}_{ij}(\hat{s},m^{2}_{Q},\mu^{2}_{F},\mu^{2}_{R})   & = & \frac{\alpha^{2}_{s}(\mu_{R})}{m^{2}_{Q}}\sum_{k=0}^{\infty}\left [ 4\pi\alpha_{s}(\mu_{R})\right ] ^{k} \sum^{a}_{l=0}f^{(k,l)}_{ij}(\rho)\ln^{l}\left ( \frac{\mu^{2}_{F}}{m^{2}_{Q}}\right )\,,
\label{equacao1}
\end{eqnarray}
where $\rho=\frac{\hat{s}}{4m^{2}_{Q}-s_0}$, $i,j=q,\bar{q},g$, specifying the types of the annihilating partons, $\hat{s}$ is the partonic center of mass, $m_{Q}$ is the heavy quark mass, $\mu_{R}$ is the renormalization scale ($s_0=1$ GeV$^2$). It is calculated as an expansion in powers of $\alpha_{s}$ with $k=0$ corresponding to the Born cross section at order ${\cal O}(\alpha^{2}_{s})$. The first correction, $k=1$, corresponds to the NLO cross section at ${\cal O}(\alpha^{3}_{s})$. To calculate the $f_{ij}$ in perturbation theory, both renormalisation and factorisation scale of mass singularities must be performed. The subtractions required are done at the mass scale $\mu$. The running of the coupling constant $\alpha_{s}$ is determined by the renormalization group. The total hadronic cross section for the heavy quark production is obtained by convoluting the total partonic cross section with the parton distribution functions of the initial hadrons \cite{nason}
\begin{eqnarray}\nonumber
\sigma_{pp}(s,m^{2}_{Q}) & = & \sum_{i,j}\int^{1}_{\tau}dx_{1}\int^{1}_{\frac{\tau}{x_{1}}}dx_{2}f^{p}_{i}(x_{1},\mu^{2}_{F})f^{p}_{j}(x_{2},\mu^{2}_{F})\hat{\sigma}_{ij}(\hat{s},m^{2}_{Q},\mu^{2}_{F},\mu^{2}_{R}),
\label{Matrix}
\end{eqnarray}
with the sum $i,j$ over all massless partons. Here, $x_{1,2}$ are the hadron momentum fractions carried by the interacting partons, $f^{p}_{i(j)}$ is the parton distribution functions, evaluated at the factorization scale and assumed to be equal to the renormalization scale in our calculations. Here, the cross sections were calculated with the following mass and scale parameters: $\mu_{c}=2m_{c}$, $m_{c}=1.5$ GeV, $\mu_{b}=m_{b}=4.5$ GeV, based on the current phenomenology for heavy quark hadroproduction \cite{15magno}.

For diffractive processes, we rely on the resolved pomeron model  where
the Pomeron structure (quark and gluon content) is probed. In the case of single diffraction, a Pomeron is emitted by one of the colliding hadrons. That hadron is detected, at least in principle, in the final state and the remaining hadron scatters off the emitted Pomeron. A typical single diffractive reaction is given by $p+p\rightarrow p+Q\bar{Q}+X$, with the cross section assumed to factorise into the total Pomeron--hadron cross section and the  Pomeron  flux  factor  \cite{IS}, $f_{{\rm\pom}/i}(x^{(i)}_{\pom},|t_i|)$.  As usual, the Pomeron kinematical variable $x_{\pom}$ is defined as $x_{\pom}^{(i)}=s_{\pom}^{(j)}/s_{ij}$, where  $\sqrt{s_{\pom}^{(j)}}$ is the center-of-mass energy in the Pomeron--hadron $j$ system and $\sqrt{s_{ij}}=\sqrt{s}$ the center-of-mass energy in the hadron $i$--hadron $j$ system. The momentum transfer in the  hadron $i$ vertex is denoted by $t_i$. A similar approach can also be applied to double Pomeron exchange (DPE) process, where both colliding hadrons  can in  principle  be detected in the final  state.  Thus, a typical reaction would be  $p+p \rightarrow p+ Q\bar{Q} + X+p$, and DPE  events are characterized by two  quasi--elastic hadrons with  rapidity  gaps between them and the central heavy flavor products. The inclusive DPE cross section may then be written as,
%%%%%%%%%%%%%%%%%%%%%%%%%%%%%%%%%%%%%%%%%%%%%%%%%%%%%%%%%%%%%%%%%%%%%%
\begin{eqnarray}
 \label{sddexp}
\frac{d\sigma(pp\rightarrow pp+ Q\bar{Q}+X)}
{dx^{(1)}_{\pom}dx^{(2)}_{\pom}d|t_1|d|t_2|}\!\! & = & \!\! f_{{\rm\pom}/p}(x^{(1)}_{\pom},|t_1|)\, f_{{\rm\pom}/p}(x^{(2)}_{\pom},|t_2|)\sum_{i,j=q,g}\!\sigma\left({\pom} + {\pom}\rightarrow  Q\bar{Q}  +  X\right), \nonumber \\
& & 
\end{eqnarray}
%%%%%%%%%%%%%%%%%%%%%%%%%%%%%%%%%%%%%%%%%%%%%%%%%%%%%%%%%%%%%%%%%%%%%%
where the Pomeron-Pomeron cross section is given by,
\begin{eqnarray}
\label{ddxsect}
\sigma\left({\pom} + {\pom}\rightarrow  Q\bar{Q}  +  X\right) &=& \int\int dx_1 \,dx_2 \,\hat \sigma_{ij}(\hat{s} ,m_Q^2,\mu^2)f_{i/\pom}\left(\beta_1,\mu^2\right) f_{j/\pom}\left(\beta_2,\mu^2\right),\nonumber \\
&  &
\end{eqnarray}
where $f_{i/\pom}\left(\beta,\mu^2\right)$ are the diffractive parton (quark, gluon) distribution functions (DPDFs) evaluated for parton   momentum fraction $\beta_{a}=x_{a}/x_{\pom}^{a}$ ($a=1,2$) and evolution scale $\mu^2$.

We further correct Eq. (\ref{sddexp})  by considering the suppression of the hard diffractive cross section by multiple-Pomeron scattering effects.  As a baseline value, we follow  Ref. \cite{KKMR}. For the present purpose, we consider $S_{\mathrm{gap}}^2=0.032\,(0.031)$ at $\sqrt{s}=5.5\, (6.3)$ TeV in nucleon-nucleon collisions, which is obtained using  a parametric interpolation formula for the KMR survival probability factor \cite{KKMR} in the form $ S_{\mathrm{gap}}^2  =a/[b+\ln (\sqrt{s/s_0})]$ with $a = 0.126$, $b=-4.688$ and $s_0=1$ GeV$^2$. This formula interpolates between survival probabilities for central diffraction (CD) in  proton-proton collisions of $4.5 \, \%$ at Tevatron and $2.6 \, \%$ at the LHC. In addition, in order to  analyze the model dependence of the cross section, we consider another approach to inclusive diffractive production of heavy quarks. In order to do so, the  Bialas-Landshoff (BL) approach \cite{Land-Nacht,Bial-Land} for the process $p+p\rightarrow p+Q\bar{Q}+p$ is taken into account.  The calculation that follows concerns  central inclusive process, where the QCD radiation accompanying the produced object is allowed. Thus, we did not include a Sudakov survival factor $T(\kappa,\mu)$ \cite{Khoze} which is needed for exclusive central processes. The cross-section is given by \cite{Bial-Szer}:
\begin{equation}
\sigma_{\pom \pom}(\mathrm{BL})=\frac{1}{2s\left(  2\pi\right)  ^{8}}\int\overline{|M_{fi}|^{2}%
}\left[  F\left(  t_{1}\right)  F\left(  t_{2}\right)  \right]  ^{2}dPH,
\label{cro-sec-ogolny}%
\end{equation}
where $F\left(  t\right)  $ is the nucleon form-factor approximated by $F\left(  t\right)  =\exp\left(b\, t\right)$, with slope parameter $b=$ $2$ GeV$^{-2}$. The differential phase-space factor is denoted by $dPH$. Following
\cite{Bial-Szer}, the use of Sudakov parameterization for momenta is given by
\begin{eqnarray}
Q  &  = &\frac{x}{s}p_{1}+\frac{y}{s}p_{2}+v,\hspace{0.5cm}k_{1}    = x_{1}p_{1}+\frac{y_{1}}{s}p_{2}+v_{1},\nonumber\\
k_{2}  &  = & \frac{x_{2}}{s}p_{1}+y_{2}p_{2}+v_{2},\hspace{0.5cm} r_{2}    =  x_{Q}p_{1}+y_{Q}p_{2}+v_{Q}, \nonumber
\end{eqnarray}
where $v,$ $v_{1},$ $v_{2},$ $v_{Q}$ are two-dimensional four-vectors describing the transverse components of the momenta. The momenta for the incoming (outgoing) protons are $p_1,\,p_2$ ($k_1,\,k_2$) and the momentum for the produced quark (antiquark) is $r_2$ ($r_1$), whereas the momentum for one of the exchanged gluons is $Q$. The square of the invariant matrix element averaged over initial spins and summed over final spins is given by \cite{Bial-Szer},
\begin{eqnarray}
\overline{\left|  M_{fi}\right|  ^{2}} & = & \frac{x_1y_2\,H}{\left(sx_{Q}y_{Q}\right)
^{2}\left(  \delta_{1}\delta_{2}\right)  ^{1+2\epsilon}\delta_{1}^{2\alpha
^{\prime}t_{1}}\delta_{2}^{2\alpha^{\prime}t_{2}}} \left(1-\frac{4\,m_Q^2}{s\delta_1\delta_2}\right)\,\exp\left[  2\beta\left(
t_{1}+t_{2}\right)  \right]  .\nonumber \\
\end{eqnarray}

In the expression above, $\delta_{1}=1-x_1$,  $\delta_{2}=1-y_2$, $t_{1}=-\vec{v}_{1}^{2}$ and $t_{2}=-\vec{v}_{2}^{2}$. The factor $\exp\left[ 2\beta\left(  t_{1}+t_{2}\right)  \right]$ takes into account the effect of the momentum transfer dependence of the non-perturbative gluon propagator with $\beta=1$ GeV$^{-2}$. The overall normalization can be expressed as,
\begin{eqnarray}
H = S_{\mathrm{gap}}^2\times2s\,\left[ \frac{4\pi m_Q\,(G^2D_0)^3\mu^4}{9\,(2\pi)^2}  \right]^2\,\left(\frac{\alpha_s}{\alpha_{0}}\right)^2,
\end{eqnarray}
where $\alpha_s$ is the perturbative coupling constant (it depends on the hard scale) and $\alpha_0$ (supposed to be independent of the hard scale) is the unknown nonperturbative coupling constant. In the numerical calculation, we use the parameters \cite{Bial-Szer} $\epsilon=0.08,$ $\alpha^{\prime}=0.25$ GeV$^{-2}$, $\mu=1.1$ GeV and $G^{2}D_{0}=30$ GeV$^{-1}\mu^{-1}$. The Regge Pomeron trajectory is then $\alpha_{\pom}(t)= 1+\epsilon+\alpha^{\prime} t$. It is taken $k_{\mathrm{min}} = 0$ for the minimum value for the transverse momentum of the quark. For the strong coupling constant, we use $\alpha_s=0.2 \,(0.17)$ for charm (bottom). An indirect determination of the unknown parameter $\alpha_0$ has been found in Ref. \cite{Adam} using experimental data for central inclusive dijet production cross section at Tevatron. Namely,  it has been found the constraint $S_{\mathrm{gap}}^2\,(\sqrt{s}=2\,\mathrm{TeV})/\alpha_0^2 = 0.6$, where $S_{\mathrm{gap}}^2$ is the gap survival probability factor (absorption factor). Considering the KMR \cite{KKMR} value $S_{\mathrm{gap}}^2=0.045$ for CD processes at Tevatron energy, one obtains $\alpha_0^2=0.075$.

\begin{table}[t]
\centering
\renewcommand{\arraystretch}{1.5}
\begin{tabular}{c c c c}
\hline
	$Q\bar{Q}$       &   $\sigma_{\mathrm{inc}}$ $[\mu b]$    &	$\sigma_{\mathrm{DPE}}$ $[\mu b]$ 	&	$R_{\mathrm{DPE}}$ [\%] \\\hline
$c\bar{c}$	&	7811           &  13.6--0.53      	&	0.17--7$\times10^{-3}$      \\
$b\bar{b}$	&	393      	    &	0.053--0.027         	&	0.01--0.007     \\
\hline
\end{tabular}
\caption{The inclusive and DPE (corrected by absorption effects) cross sections in $pp$ collisions at the LHC (14 TeV). For the inclusive diffractive cross section the first value corresponds to the resolved pomeron mode and the second one BL model. The corresponding diffractive ratios, $R_{\mathrm{DPE}}$, are also presented.}
\label{tabelahadronica}
\end{table}

 The calculations for the inclusive and diffractive cross sections as well as the diffractive ratios to heavy quark production in proton-proton collisions are showed at Tab. (\ref{tabelahadronica}). For the inclusive diffractive cross section the first value corresponds to the partonic picture of Pomeron,  Eq. (\ref{sddexp}), and the second one to the BL approach, Eq. (\ref{cro-sec-ogolny}). We assume the value $S_{\mathrm{gap}}^2=0.026$ for the absorption corrections at energy of 14  TeV. The partonic PDFs and scales are mentioned in previous section. For the diffractive gluon PDF, we take the experimental (H1 collaboration) FIT A \cite{H1FPS}. The main theoretical uncertainty in the diffractive ratio is the survival probability factor, whereas uncertainties associated to factorization/renormalization scale, parton PDFs and quark mass are minimized taking a ratio. The present results are consistent with a previous estimate performed in Ref. \cite{QQmvtm}, where a value $S_{\mathrm{gap}}^2=0.04$ was considered and cross sections were computed at LO accuracy.

The single diffraction case can be also addresses, where the reaction is given by $p+p\rightarrow p+Q\bar{Q}+X$.  The single diffractive cress section may then be written as \cite{QQmvtm}
\begin{eqnarray}
 \label{sdexp}
\frac{d\sigma^{\mathrm{SD}}\,(pp\rightarrow p+  Q\bar{Q}+X)}
{dx^{(a)}_{\pom}d|t_a|}\! & = & \! f_{{\rm\pom}/a}(x^{(a)}_{\pom},|t_a|)\sigma\left({\pom} + p\rightarrow  Q\bar{Q}  +  X\right),
\end{eqnarray}
where $x_{\pom}$ is the Pomeron kinematical variable, defined as $x^{a}_{\pom}=s^{(b)}_{\pom}/s_{ab}$, where $\sqrt{s^{b}_{\pom}}$ is the center-of-mass energy in the Pomeron-hadron $b$ system and $\sqrt{s}_{ab}=\sqrt{s}$ is the center-of-mass energy in the hadron$_{a}$-hadron$_{b}$ system, with $t_{a}$ denoting the momentum transfer in the hadron $a$ vertex. In terms of pomeron pdfs and pomeron flux, the expression for the single diffractive cross section for $Q\bar{Q}$ production is written as \cite{QQmvtm}
\begin{eqnarray}
\label{sdxsect}
& & \sigma_{ab}^{\mathrm{SD}}(s,m_Q^2)  =   \sum_{i,j=q\bar{q},g}
\int_{\rho}^1dx_1\int_{\rho/x_1}^1 dx_2 \int_{x_1}^{x_{\pom}^{\mathrm{max}}}\frac{dx_{\pom}^{(1)}}{x_{\pom}^{(1)}} \nonumber \\ \nonumber & \times & \bar{f}_{\pom/a}\left(x_{\pom}^{(1)}\right)f_{i/\pom}\left(\frac{x_1}{x_{\pom}^{(1)}},\mu^2\right)  f_{j/b}(x_2,\mu^2)\,
\hat \sigma_{ij}(\hat{s} ,m_Q^2,\mu^2) \, (1\rightleftharpoons 2).
\end{eqnarray}

The calculations for the inclusive and diffractive cross sections, as well the diffractive ratios to heavy quark production in proton-proton collisions are showed at Tab. (\ref{tabelahadronicasd}). We take the value $S_{\mathrm{gap}}^2=0.06$ for the absorption corrections in hadronic collisions at the LHC. The partons PDF and scales are mentioned in previous section. For the diffractive gluon PDF, we take the experimental FIT A (the fully integrated cross section is insensitive to a different choice, i.e. FIT B). The main theoretical uncertainty in the diffractive ratio is the survival probability factor, whereas uncertainties associated to factorization/renormalization scale, parton PDFs and quark mass are minimized taking a ratio.

\begin{table}[t]
\centering
\renewcommand{\arraystretch}{1.5}
\begin{tabular}{cccc}
\hline
	Heavy Quark       &   $\sigma_{\mathrm{inc}}$ $[\mu b]$   &	$\sigma_{\mathrm{SD}}$  $[\mu b]$	&	$R_{\mathrm{SD}}$ \\\hline
$c\bar{c}$	&	7811           &  		178     	&		2.3 \%      \\ \hline 
$b\bar{b}$	&	393      	    &	    	7        	&		1.7	\%    \\\hline
\end{tabular}
\caption{The inclusive and single diffractive (corrected by absorption effects) cross sections in $pp$ collisions at the LHC (14 TeV). The corresponding diffractive ratios, $R_{\mathrm{SD}}$, are also presented.}
\label{tabelahadronicasd}
\end{table}

As a last example, we consider the diffractive quarkonium production at the LHC.  We use the Color Evaporation Model (CEM) \cite{CEM} for the production model. The main reasons for this choice  are its simplicity and fast phenomenological implementation, which are the base for its relative success in describing high energy data.  In this model, the cross section for a process in which partons of two hadrons, $h_1$ and $h_2$, interact to produce a heavy quarkonium state, $h_1 + h_2 \rightarrow H(nJ^{\mathrm{CP}}) + X$, is given by the cross section of open heavy-quark pair production that is summed over all spin and color states. All information on the non-perturbative transition of the $Q\bar{Q}$ pair to the heavy quarkonium $H$ of quantum numbers $J^{\mathrm{PC}}$ is contained in the factor $F_{nJ^{\mathrm{PC}}}$ that {\it a priori} depends on all quantum numbers \cite{CEM},
\begin{eqnarray}
\sigma (h_1\,h_2 \rightarrow H[nJ^{\mathrm{CP}}]\,X)= F_{nJ^{\mathrm{PC}}}\,\bar{\sigma}(h_1 \,h_2 \rightarrow Q\bar{Q}\, X)\,,
\end{eqnarray}
where $\bar{\sigma}(Q\bar{Q})$ is the total hidden cross section of open heavy-quark production calculated by integrating over the $Q\bar{Q}$ pair mass from $2m_Q$ to $2m_O$, with $m_O$ is the mass of the associated open meson. The hidden cross section can be obtained from the usual expression for the total
cross section to NLO as mentioned before. Here, we assume that the factorization scale, $\mu_F$, and the renormalization scale, $\mu_R$, are equal, $\mu = \mu_F = \mu_R$. We also take $\mu=2m_Q$, using the quark masses $m_c=1.2$ GeV and $m_b=4.75$ GeV. These parameters provide an adequate description of open heavy-flavour production \cite{Mangano}. The invariant mass is integrated over $4m_c^2\leq \hat{s}\leq 4m_D^2$ in the charmonium case and  $4m_b^2\leq \hat{s}\leq 4m_B^2$ for $\Upsilon$ production. The factors $F_{nJ^{\mathrm{PC}}}$ are experimentally determined \cite{MQD} to be $F_{11^{--}}\approx 2.5\times 10^{-2}$ for $J/\Psi$ and $F_{11^{--}}\approx 4.6\times 10^{-2}$ for $\Upsilon$. These coefficients are obtained with NLO cross sections for heavy quark production \cite{MQD}.

In the resolved pomeron model, the diffractive process  $p+p\rightarrow pp+H[nJ^{\mathrm{CP}}]+X$,  may then be written as
%%%%%%%%%%%%%%%%%%%%%%%%%%%%%%%%%%%%%%%%%%%%%%%%%%%%%%%%%%%%%%%%%%%%%%
\begin{eqnarray}
 \label{sdexp}
 \frac{d\sigma^{\mathrm{SD}}\,(p_i+p_j\rightarrow p_{i}+  H[nJ^{\mathrm{CP}}]+X)}{dx^{(i)}_{\pom}d|t_i|}   & = & F_{nJ^{\mathrm{PC}}}f_{{\rm\pom}/p_i}(x^{(i)}_{\pom},|t_i|)\, \bar{\sigma}\left({\pom} + p_{j}\rightarrow  Q\bar{Q}  +  X\right), \nonumber\\
& & 
\end{eqnarray}
%%%%%%%%%%%%%%%%%%%%%%%%%%%%%%%%%%%%%%%%%%%%%%%%%%%%%%%%%%%%%%%%%%%%%%
where the Pomeron kinematical variable $x_{\pom}$ is defined as $x_{\pom}^{(i)}=s_{\pom}^{(j)}/s_{ij}$, where  $\sqrt{s_{\pom}^{(j)}}$ is the center-of-mass energy in the Pomeron--hadron $j$ system and $\sqrt{s_{ij}}=\sqrt{s}$ the center-of-mass energy in the hadron $i$--hadron $j$ system. The momentum transfer in the  hadron $i$ vertex is denoted by $t_i$. A similar factorization can also be applied to central diffraction, where both colliding
hadrons  can in  principle  be detected in the final  state. The central quarkonium production, $p_1+p_2 \rightarrow p_1+
 H[nJ^{\mathrm{CP}}] + p_2$, is characterized by
two  quasi--elastic hadrons with  rapidity  gaps between them and the
central heavy quarkonium products. The double pomeron exchange cross section  may then be written as,
%%%%%%%%%%%%%%%%%%%%%%%%%%%%%%%%%%%%%%%%%%%%%%%%%%%%%%%%%%%%%%%%%%%%%%
\begin{eqnarray}
 \label{sddexp}
& & \frac{d\sigma^{\mathrm{DPE}}\,(p_i+p_j\rightarrow p_i + H[nJ^{\mathrm{CP}}]   +p_j)}
{dx^{(i)}_{\pom}dx^{(j)}_{\pom}d|t_i|d|t_j|} =  F_{nJ^{\mathrm{PC}}} \times\nonumber \\
& &  f_{{\rm\pom}/i}(x^{(i)}_{\pom},|t_i|)\, f_{{\rm\pom}/j}(x^{(j)}_{\pom},|t_j|) \,\bar{\sigma}\left({\pom} + {\pom}\rightarrow  Q\bar{Q}  +  X\right).\nonumber
\end{eqnarray}
%%%%%%%%%%%%%%%%%%%%%%%%%%%%%%%%%%%%%%%%%%%%%%%%%%%%%%%%%%%%%%%%%%%%%%

Here, we assume that one of the hadrons, say proton $p_1$, emits a Pomeron whose partons interact with partons of the proton $p_2$. Using the same notation as for the diffractive heavy quark production, the hidden heavy flavour cross section can be obtained from Pomeron-hadron cross sections for  single  and  central diffraction processes,
%%%%%%%%%%%%%%%%%%%%%%%%%%%%%%%%%%%%%%%%%%%%%%%%%%%%%%%%%%%%%%%%%%%%%%
\begin{eqnarray}
\label{sdxsect}
& &\frac{d\sigma\left({\pom} + h \rightarrow  Q\bar{Q}  +  X\right)}{dx_1\,dx_2}     =   \sum_{i,j=q\bar{q},g} \frac{f_{i/\pom}\left(x_1/x_{\pom}^{(1)};\,\mu^2_F\right)}{x_{\pom}^{(1)}} \nonumber \\
&\times & f_{j/h_2}(x_2,\mu^2_F)\,
\hat \sigma_{ij}(\hat{s} ,m_Q^2,\mu^2_F,\mu^2_R) +\, (1\rightleftharpoons 2)\,,
\end{eqnarray}

and

%%%%%%%%%%%%%%%%%%%%%%%%%%%%%%%%%%%%%%%%%%%%%%%%%%%%%%%%%%%%%%%%%%%%%%
\begin{eqnarray}
\label{ddxsect}
& &\frac{d\sigma\left({\pom} + {\pom} \rightarrow  Q\bar{Q}  +  X\right)}{dx_1\,dx_2}     =  \sum_{i,j=q\bar{q},g} \frac{ f_{i/\pom}\left(x_1/x_{\pom}^{(1)};\,\mu^2_F\right)}{x_{\pom}^{(1)}}\,\nonumber \\
& & \times \frac{f_{j/\pom}\left(x_2/x_{\pom}^{(2)};\,\mu^2_F\right) }{x_{\pom}^{(2)}} \hat \sigma_{ij}(\hat{s} ,m_Q^2,\mu^2_F,\mu^2_R).\nonumber
\end{eqnarray}

%%%%%%%%%%%%%%%%%%%%%%%%%%%%%%%%%%%%%%%%%%%%%%%%%%%%%%%%%%%%%

We will consider the theoretical estimates for $S_{\mathrm{gap}}^2$ from Ref. \cite{KKMR}, which considers a two-channel eikonal model and rescattering effects.  Thus, we have $S_{\mathrm{gap}}^2(SD)=0.15,\,[0.09]$ and  $S_{\mathrm{gap}}^2(DPE)=0.08,\,[0.04]$ for $\sqrt{s}=1.8$ TeV (Tevatron) [$\sqrt{s}=14$ TeV (LHC)].

\begin{table}[t]
\centering
\renewcommand{\arraystretch}{1.5}
\begin{tabular}{cccc}
$\sqrt{s}$  & Quarkonium   & $R_{\mathrm{SD}}$ (\%) & $R_{\mathrm{DPE}}$ (\%)\\
\hline
2.0 TeV &  $J/\Psi$  & $0.93 $   & $ 0.2$\\
14 TeV &  $J/\Psi$  & $0.50$  & $0.15$\\
2.0 TeV &    $(\Upsilon+\Upsilon^{\prime}+\Upsilon^{\prime \prime})$  & $0.78$  & $0.06$\\
14 TeV &     $(\Upsilon+\Upsilon^{\prime}+\Upsilon^{\prime \prime})$  & $0.39$   & $0.03$\\
\hline
\end{tabular}
\caption{\label{tab:table1} Model predictions for single and DPE diffractive  quarkonium  production \cite{MQD} in Tevatron and the LHC (14 TeV). }
\end{table}
%%%%%%%%%%%%%%%%%%

%%%%%%%%%%%%%%%%%%%%%%%%%%%%%%%%%%%%%%%%%%%%%%%%%%%%%%%%%%%%%

\section{Exclusive diffractive processes in $ep$ collisions}
\label{exclusive}

Let us now consider the case of  diffractive processes where a photon dissociates into a single particle.  Due to the vacuum quantum numbers carried by the pomeron this particle can in particular be a vector meson having the same photon quantum
numbers. In addition, we can have also a deeply virtual Compton scattering (DVCS), $\gamma^*p \to \gamma
p$. It is known from HERA data that the energy dependence of these processes
becomes steep in the presence of a hard scale, which can be either the
photon virtuality $Q^2$ or the mass of the meson in the case of
$J/\Psi$ or $\Upsilon$ production \cite{rp8}.  This is similar to the energy
dependence of the $\gamma^* p$ total cross section, which changes from flat to steep when going from real photons to
$Q^2$ of a few GeV$^2$.  In pQCD diffraction proceeds by two-gluon exchange and the transition from a
virtual photon to a real photon or to a $q\bar{q}$-pair
subsequently hadronizing into a meson is a short-distance process
involving these gluons,
provided that either $Q^2$ or the quark mass is large.  Within certain approximations, the DVCS and vector
meson cross sections are proportional to the square of the gluon distribution in
the proton, evaluated at a scale of order $Q^2+M_V^2$ and at a momentum
fraction $\xpom= (Q^2+M_V^2) /(W^2+Q^2)$, where the vector meson mass
$M_V$ now takes the role of $M_X$ in inclusive diffraction
\cite{Ryskin:1992ui}. In analogy to the case of the total $\gamma^* p$
cross section,
the energy dependence of the cross sections  thus reflects the $x$
and scale dependence of the gluon density in the proton, which grows
with decreasing $x$ with a slope becoming steeper as the scale
increases.

The inclusive DIS cross section is related to the imaginary part of the forward virtual Compton
amplitude. Therefore, the usual gluon distribution gives the probability to find one gluon
in the proton. On the other hand, the corresponding graphs for DVCS and exclusive meson
 production  represent the amplitudes of
 exclusive processes, which are proportional to the probability
 amplitude for first extracting a gluon from the initial proton and then
 returning it to form the proton in the final state.  Making use of high energy theorems, at large $Q^2$ the short-distance factorization
holds, in analogy to the case of inclusive DIS.  QCD factorization
theorems~\cite{Collins:1996fb} state that in the limit of large $Q^2$ the Compton amplitude factorizes into a hard-scattering
subprocess and a hadronic matrix element describing the emission and
reabsorption of a parton by the proton target.  The
analogous result for exclusive meson production involves in addition
the $q\bar{q}$ distribution amplitude of the meson, the so-called meson wave function, which is a non-perturbative
input. In this collinear factorization approach, the hadronic matrix elements for exclusive processes are not the usual PDFs as the proton has not the same momentum in the initial and final state. Consequently, they are more general functions named generalized PDFs taking into account the momentum difference between the initial and final state proton.

In last decades,  a different type of factorization has been very fruitful in phenomenology. It is the high-energy or $k_t$ factorization approach, which is based on the BFKL formalism.  Now, the gluon distribution appearing in the factorization formulae depends
explicitly on the transverse momentum $k_t$ of the emitted gluon.  In
collinear factorization, this $k_t$ is integrated over in the parton
distributions and set to zero when calculating the hard-scattering
process.  Similarly, the meson wave functions appearing in
$k_t$ factorization explicitly depend on the relative transverse
momentum between the $q$ and $\bar{q}$ in the meson, whereas this is
integrated over in the quark-antiquark distribution amplitudes of the collinear approach.  The two formalisms implement different ways of
separating different parts of the dynamics in a scattering process.
The building blocks in a short-distance factorization formula correspond
to
  either small or large particle virtuality, whereas the separation criterion
   in high-energy
factorization is the particle rapidity.

The different building blocks in the graphs for Compton scattering and
meson production  can be rearranged in a such way that it admits a
very intuitive interpretation in a reference frame where the photon
carries large momentum. In the typical proton rest frame the initial photon splits into a
quark-antiquark pair, which scatters on the proton and finally forms a
photon or meson again.  In addition, one can perform a Fourier transformation and trade the
relative transverse momentum between quark and antiquark for their
transverse distance $r$, which is conserved in the scattering on the
target.  The quark-antiquark pair acts as a color dipole, and its
scattering on the proton is described by a {\it dipole cross section},
$\sigma_{q\bar{q}}$ depending on $r$ and on $\xpom$ (or on $x$ in the
case of inclusive DIS).  The wave functions of the photon and the meson
depend on $r$ after Fourier transformation, and at small $r$ the photon
wave function is perturbatively calculable.  Typical values of $r$ in a
scattering process are determined by the inverse of the hard momentum
scale, i.e. $r \sim (Q^2+M_V^2)^{-1/2}$.  An important result of
high-energy factorization is the relation
\begin{equation}
  \label{dip-gluon}
\sigma_{q\bar{q}}(r, x) = \frac{\pi^2}{3} \,\alpha_S(A/r) \,r^2 x g(x,A/r),
\end{equation}
at small $r$, where we have replaced the generalized gluon
distribution by the usual
one in the spirit of the leading $\log x$ approximation.  A more
precise version of the relation (\ref{dip-gluon}) involves the $k_t$
dependent gluon distribution.  The dipole cross section
vanishes at $r=0$ in accordance with the phenomenon of {\it color
transparency}, where a hadron becomes more and more transparent for a color
dipole of decreasing size.

\subsection{The parton saturation phenomenom and saturation models}

Diffraction involves scattering on
small-$x$ gluons in the proton.  Taking the density in the
transverse plane of gluons with longitudinal momentum fraction $x$
that are resolved in a process with hard scale $Q^2$ one can think
of $1/Q$ as the {\it transverse size} of these gluons as seen by the
probe.  The number density of gluons at given $x$ increases with
increasing $Q^2$, as described by DGLAP evolution.  According to the BFKL evolution equation it
also increases at given $Q^2$ when $x$ becomes smaller, so that the
gluons become more and more densely packed.  At some point, they will
start to overlap and thus reinteract and screen each other.  One then
enters a regime where the density of partons saturates and where the
linear DGLAP and BFKL evolution equations cease to be valid.  If $Q^2$
is large enough to have a small coupling $\alpha_s$, we have a theory
of this non-linear regime called {\it Color Glass Condensate} (CGC) \cite{rp4,rp5}.  To quantify the onset of non-linear
effects, one introduces a saturation scale $Q^2_s$ depending on $x$,
such that for $Q^2 < Q^2_s(x)$ these effects become important.  For
smaller values of $x$, the parton density in the target proton is
higher, and saturation sets in at larger values of $Q^2$.

The color dipole picture  is well suited for the theoretical description of saturation effects.  When
such effects are important, the relation (\ref{dip-gluon}) between
dipole cross section and gluon distribution ceases to be valid; in
fact the gluon distribution itself is then no longer an adequate
quantity to describe the dynamics of a scattering process.  In a
certain approximation, the evolution of the dipole cross section with
$x$ is described by the Balitsky-Kovchegov
equation~\cite{Balitsky:1995ub}, which supplements the BFKL equation
with a non-linear term taming the growth of the dipole cross section
with decreasing $x$. Essential features of the saturation phenomenon are captured in a
phenomenological model for the dipole cross section, originally
proposed by Golec-Biernat and W\"usthoff, see
\cite{Golec-Biernat:1999qd,Golec-Biernat:2002bj}. In this model, the dipole size $r$ now plays the role of
$1/Q$.  At small $r$ the cross section rises
following the relation $\sigma_{q\bar{q}}(r, x) \propto r^2 x g(x)$.
At some value $R_s(x)$ of $r$, the dipole cross section is so large
  that this relation ceases to be valid, and $\sigma_{q\bar{q}}$ starts to
  deviate from the quadratic behavior in $r$.
  As $r$ continues to increase, $\sigma_{q\bar{q}}$ eventually saturates
  at a value typical of a meson-proton cross section.
In terms of the
saturation scale introduced above, $R_s(x) = 1/Q_s(x)$.  For
smaller values of $x$, the initial growth of $\sigma_{q\bar{q}}$ with $r$
is stronger because the gluon distribution is larger.  The target is thus
more opaque and as a consequence saturation sets in at lower $r$.

An important feature found both in this phenomenological
model~\cite{Stasto:2000er} and in the solutions of the
Balitsky-Kovchegov equation \cite{Golec-Biernat:2001if} is
that the total $\gamma^* p$ cross section only depends on $Q^2$ and
$x_B$ through a single variable $\tau= Q^2 /Q^2_s(x)$.  This
property, referred to as {\it geometric scaling}, is well satisfied by the
data at small $x_B$ and is an important
piece of evidence that saturation effects are visible in these data.
Phenomenological estimates find $Q^2_s$ of the order 1~GeV$^2$ for
$x$ around $10^{-3}$ to $10^{-4}$. The dipole formulation is suitable to describe not only exclusive
  processes and inclusive DIS, but also inclusive diffraction
  $\gamma^* p\to X p$.
For a diffractive final state $X=q\bar{q}$ at
parton level, the theory description is very similar to the one for
deeply virtual Compton scattering, with the wave function for the
final state photon replaced by plane waves for the produced $q\bar{q}$
pair. The inclusion of the case $X=q\bar{q}g$ requires further
approximations~\cite{Golec-Biernat:1999qd} but is phenomenologically
indispensable for moderate to small $\beta$. In next section, we give some examples where one uses the phenomenological models including saturation physics to compute the exclusive production of particles (vector mesons or real photons at the final state) and compare them to the available high precision HERA data.

\section{Examples of phenomenology in photon-proton interactions}
\label{samplesphpd}
%------------------------------------------------------------------------------
Let us consider photon-hadron scattering in the color dipole frame, in which most of the energy is
carried by the hadron, while the  photon  has
just enough energy to dissociate into a quark-antiquark pair
before the scattering. In this representation the probing
projectile fluctuates into a
quark-antiquark pair (a dipole) with transverse separation
$\rr$ long after the interaction, which then
scatters off the hadron \cite{nik}.
In the dipole picture the   amplitude for production of an exclusive final state $E$, such as a vector meson ($E = V$) or a real photon in DVCS ($E = \gamma$) is given by (See e.g. Refs. \cite{nik,vicmag_mesons,non})
\begin{eqnarray}
\, {\cal A}_{T,L}^{\gamma^*p \rightarrow E p}\, (x,Q^2,\Delta)  =
\int dz\, d^2\rr \,(\Psi^{E*}\Psi)_{T,L}\,{\cal{A}}_{q\bar{q}}(x,\rr,\Delta) \, ,
\label{sigmatot}
\end{eqnarray}
where $(\Psi^{E*}\Psi)_{T,L}$ denotes the overlap of the photon and exclusive final state wave functions. The variable  $z$ $(1-z)$ is the
longitudinal momentum fractions of the quark (antiquark),  $\Delta$ denotes the transverse momentum lost by the outgoing proton ($t = - \Delta^2$) and $x$ is the Bjorken variable. For DVCS, the amplitude involves a sum over quark flavors. Moreover, ${\cal{A}}_{q\bar{q}}$ is the elementary elastic amplitude for the scattering of a dipole of size $\rr$ on the target. It is directly related to ${\cal{N}} (x,\rr,\rb)$ and consequently to the QCD dynamics (see below). One has that \cite{non}
\begin{eqnarray}
{\cal{A}}_{q\bar{q}} (x,\rr,\Delta) & = & i \int d^2 \rb \, e^{-i \rb.\rd}\, 2 {\cal{N}}(x,\rr,\rb) \,\,,
\end{eqnarray}
where $\rb$ is the transverse distance from the center of the target to one of the $q \bar{q}$ pair of the dipole.  Consequently, one can express the amplitude for the exclusive production of a final state $E$ as follows
\begin{eqnarray}
 {\cal A}_{T,L}^{\gamma^*p \rightarrow E p}(x,Q^2,\Delta) & = & i
\int dz \, d^2\rr \, d^2\rb  e^{-i[\rb-(1-z)\rr].\rd} (\Psi_{E}^* \Psi)_T \,2 {\cal{N}}(x,\rr,\rb),\nonumber \\
\label{sigmatot2}
\end{eqnarray}
where  the factor $[i(1-z)\rr].\rd$ in the exponential  arises when one takes into account non-forward corrections to the wave functions \cite{non}.
Finally, the differential cross section  for  exclusive production is given by
\begin{eqnarray}
\frac{d\sigma_{T,L}}{dt} (\gamma^* p \rightarrow E p) = \frac{1}{16\pi} |{\cal{A}}_{T,L}^{\gamma^*p \rightarrow E p}(x,Q^2,\Delta)|^2\,(1 + \beta^2)\,,
\label{totalcs}
\end{eqnarray}
where $\beta$ is the ratio of real to imaginary parts of the scattering
amplitude. For the case of heavy mesons, skewness corrections are quite important and they are also taken  into account. (For details, see Refs. \cite{vicmag_mesons,non}).

The photon wavefunctions appearing in Eq. (\ref{sigmatot2}) are well known in literature \cite{nik}. For the meson wavefunction, we have considered the Gauss-LC  model which is a simplification of the DGKP wavefunctions (for a review on the meson wavefunctions see Ref. \cite{rp8}). The motivation for this choice is its simplicity and the fact that the results are not sensitive to a different model. In photoproduction, this leads only to an  uncertainty  of a few percents in overall normalization. We consider the quark masses $m_{u,d,s} = 0.14$ GeV, $m_c = 1.4$ GeV and $m_b=4.5$ GeV. The parameters for the meson wavefunction can be found in Ref. \cite{vicmag_mesons}. In the DVCS case, as one has a real photon at the initial state, only the transversely polarized overlap function contributes to the cross section.  Summed over the quark helicities, for a given quark flavor $f$ it is given by,
\begin{eqnarray}
  (\Psi_{\gamma}^*\Psi)_{T}^f  =  \frac{N_c\,\alpha_{\mathrm{em}}e_f^2}{2\pi^2}\left\{\left[z^2+\bar{z}^2\right]\varepsilon_1 K_1(\varepsilon_1 r) \varepsilon_2 K_1(\varepsilon_2 r)  +  m_f^2 K_0(\varepsilon_1 r) K_0(\varepsilon_2 r)\right\}, \nonumber \\
  \label{eq:overlap_dvcs}
\end{eqnarray}
where we have defined the quantities $\varepsilon_{1,2}^2 = z\bar{z}\,Q_{1,2}^2+m_f^2$ and $\bar{z}=(1-z)$. Accordingly, the photon virtualities are $Q_1^2=Q^2$ (incoming virtual photon) and $Q_2^2=0$ (outgoing real photon).

The scattering amplitude ${\cal{N}}(x,\rr,\rb)$   contains all
information about the target and the strong interaction physics.
In the Color Glass Condensate (CGC)  formalism \cite{rp4,rp5}, it  encodes all the
information about the
non-linear and quantum effects in the hadron wave function. It can be obtained by solving an appropriate evolution
equation in the rapidity $y\equiv \ln (1/x)$, which in its  simplest form is the Balitsky-Kovchegov equation.  In leading
order (LO), and in the translational invariance approximation---in which the scattering
amplitude does not depend on the collision impact parameter $\bm{b}$---it reads

	\begin{eqnarray}\label{eq:bklo}
		\frac{\partial {\cal{N}}(r,Y)}{\partial Y}  =  \int {\rm d}\bm{r_1}\, K^{\rm{LO}}
		(\bm{r,r_1,r_2})
		[{\cal{N}}(r_1,Y)+{\cal{N}}(r_2,Y) -  {\cal{N}}(r,Y)-{\cal{N}}(r_1,Y){\cal{N}}(r_2,Y)], \nonumber \\
	\end{eqnarray}
where ${\cal{N}}(r,Y)$ is the scattering amplitude for a dipole (a quark-antiquark pair)
off a target, with transverse size $r\equiv |\bm{r}|$, $Y\equiv \ln(x_0/x)$ ($x_0$ is the value of $x$ where the evolution starts), and $\bm{r_2 = r-r_1}$. $K^{\rm{LO}}$ is the evolution kernel, given by

	\begin{equation}\label{eq:klo}
		K^{\rm{LO}}(\bm{r,r_1,r_2}) = \frac{N_c\alpha_s}{2\pi^2}\frac{r^2}{r_1^2r_2^2},
	\end{equation}
where $\alpha_s$ is the (fixed) strong coupling constant. This equation is a
generalization of the linear BFKL equation (which corresponds of the first three terms), with the inclusion
of the (non-linear) quadratic term, which damps the indefinite growth of the amplitude
with energy predicted by BFKL evolution. It has been shown \cite{mp} to be in the same
universality class of the Fisher-Kolmogorov-Pertovsky-Piscounov (FKPP) equation
\cite{fkpp} and, as a consequence, it admits the so-called traveling wave solutions.
This means that, at asymptotic rapidities, the scattering amplitude is a wavefront which
travels to larger values of $r$ as $Y$ increases, keeping its shape unchanged. Thus,
in such asymptotic regime, instead of depending separately on $r$ and $Y$, the amplitude
depends on the combined variable $rQ_s(Y)$, where $Q_s(Y)$ is the saturation scale. This property of the solution of BK equation is a natural
explanation to the {\it geometric scaling}, a phenomenological feature observed at the
DESY $ep$ collider HERA, in the measurements of inclusive and exclusive processes
\cite{scaling,marquet,prl,prl1}. Although having its properties been intensely studied and understood, both numerically
and analytically, the LO BK equation presents some difficulties when applied to study
DIS small-$x$ data. In particular, some studies concerning this equation
\cite{IANCUGEO,MT02,AB01,BRAUN03,AAMS05} have
shown that the resulting saturation scale grows much faster with increasing energy
($Q_s^2\sim x^{-\lambda}$, with $\lambda\simeq 4.88N_c\alpha_s/\pi \approx 0.5$ for  $\alpha_s = 0.2$) than that
extracted from phenomenology ($\lambda \sim 0.2-0.3$). This difficulty could be solved by
considering smaller values of the strong coupling constant $\alpha_s$, but this procedure
would lead to physically unrealistic values. One can conclude that higher order corrections
to LO BK equation should be taken into account to make it able to describe the available
small-$x$ data.

\begin{figure}[t]
\includegraphics[scale=0.4]{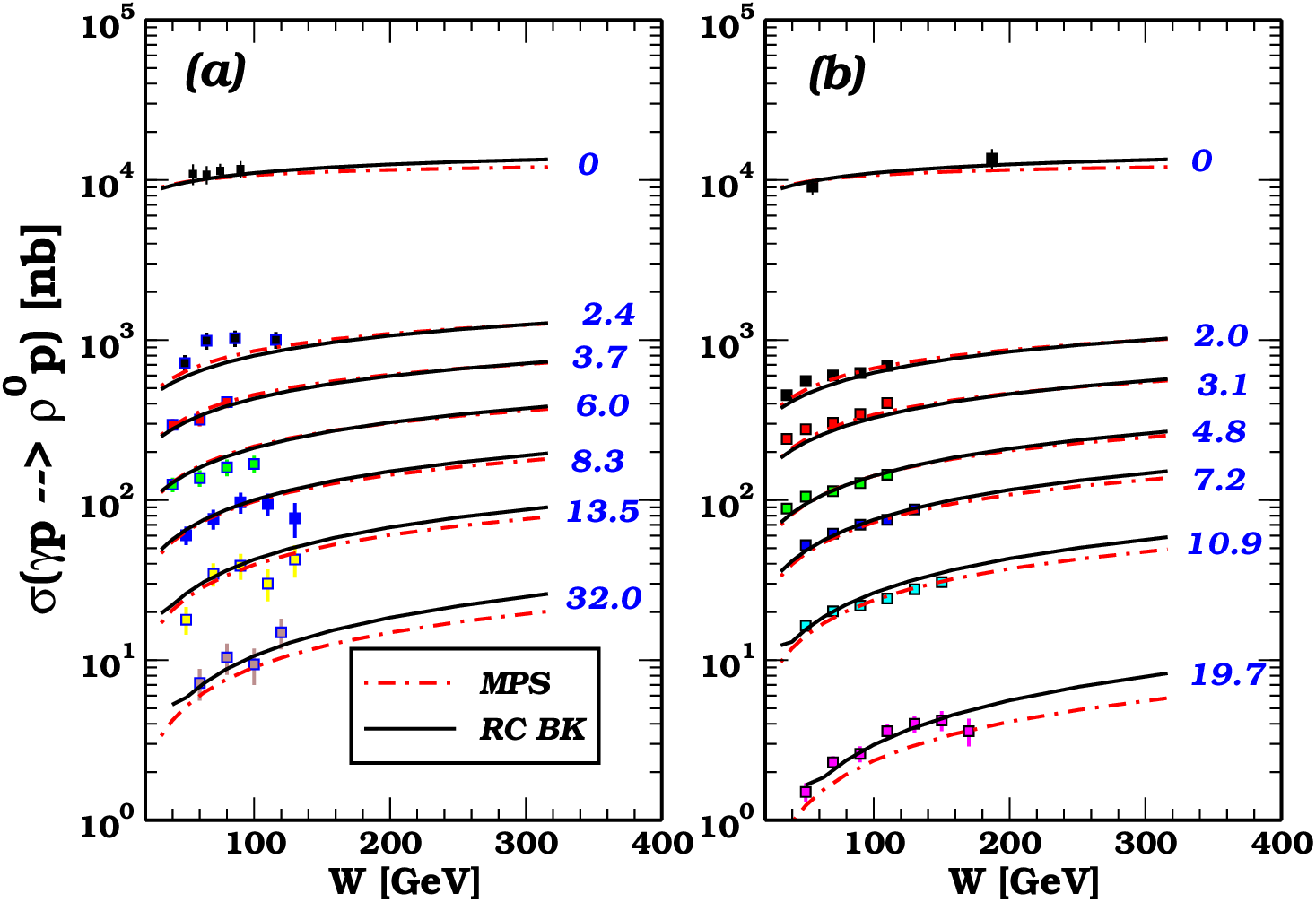}
\caption{Energy dependence of the $\gamma p$ cross section for $\rho^0$ production  for different photon virtualities. Data from (a) ZEUS and (b) H1 collaborations \cite{H1_rho,ZEUS_rho}.}
\label{fig:1}
\end{figure}

\begin{figure}[t]
\includegraphics[scale=0.4]{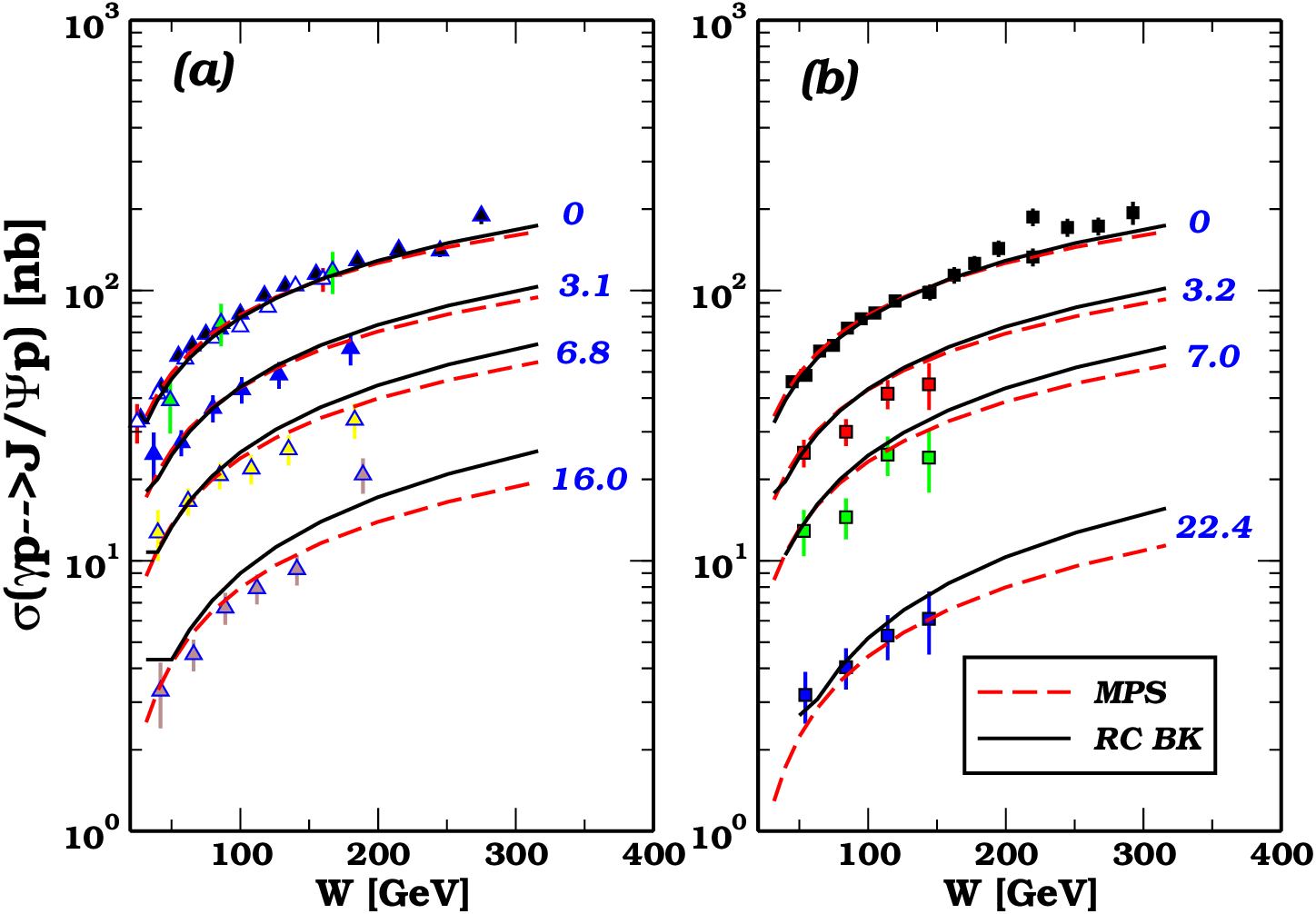}
\caption{Energy dependence of the $\gamma p$ cross section for $J/\Psi$ production  for different photon virtualities. Data from (a) ZEUS and (b) H1 collaborations \cite{ZEUS_jpsi, H1_jpsi}.}
\label{fig:2}
\end{figure}

The calculation of the running coupling corrections to BK evolution kernel was explicitly
performed in \cite{kovwei1,balnlo}, where the authors included $\alpha_sN_f$ corrections to the kernel to all orders. The  improved  BK equation is given in terms of a
running coupling and a subtraction term, with the latter accounting for conformal, non running coupling contributions. In the prescription proposed by Balitsky in \cite{balnlo} to single out the ultra-violet divergent contributions from the finite ones that originate after the resummation of quark loops, the contribution of the subtraction term is mmized at large energies. In \cite{bkrunning} this contribution was disregarded, and the improved BK equation was numerically solved replacing the leading order kernel  in Eq. (\ref{eq:bklo}) by the modified kernel which includes the running coupling
corrections and  is given by \cite{balnlo}
	\begin{eqnarray}\label{eq:krun}
		K^{\rm{Bal}}(\bm{r,r_1,r_2})  =  \frac{N_c\alpha_s(r^2)}{2\pi^2}
		\left[\frac{r^2}{r_1^2r_2^2}
              +  \frac{1}{r_1^2}\left(\frac{\alpha_s(r_1^2)}
		{\alpha_s(r_2^2)}-1\right)  +   \frac{1}{r_2^2}\left(\frac{\alpha_s(r_2^2)}
		{\alpha_s(r_1^2)}-1\right)\right] .
	\end{eqnarray}
From a recent numerical study of the improved BK equation \cite{javier_kov}, it has been confirmed that the running coupling corrections lead to a considerable
increase in the anomalous dimension and to a slow-down of the evolution
speed, which implies, for example, a slower growth of the saturation scale with
energy, in contrast with the faster growth predicted by the LO BK equation. Moreover,
as shown in \cite{bkrunning,vic_joao,alba_marquet} the improved BK equation has been shown to
be really successful when applied to the description of the $ep$ HERA data for the inclusive and diffractive proton structure function, as well as for the  forward hadron  spectra in $pp$ and $dA$ collisions.
It is important to emphasize that the impact parameter dependence was not taken into account in Ref. \cite{bkrunning}, the normalization of the dipole cross section was fitted to data and two distinct initial conditions, inspired in the Golec Biernat-Wusthoff (GBW) \cite{GBW} and McLerran-Venugopalan (MV) \cite{MV} models, were considered. The predictions resulted to be almost independent of the initial conditions and, besides, it was observed that it is impossible to describe the experimental data using only the linear limit of the BK equation, which is equivalent to Balitsky-Fadin-Kuraev-Lipatov (BFKL) equation \cite{bfkl}. In next section we will compare the results of the RC BK approach to the experimental data on exclusive processes at DESY-HERA and present our predictions for the kinematical range of a future electron - proton collider \cite{dainton}.

In what follows we calculate the exclusive observables using as input in our calculations the solution of the RC BK evolution equation. The results has been published in Ref. \cite{code}. In numerical calculations we have considered the GBW initial condition for the evolution (we quote Ref. \cite{bkrunning}  for details) and it was verified the MV initial condition gives cross section with overall normalization $10-15\,\%$ smaller and unchanged energy dependence. Furthermore, we compare the RC BK predictions with those from the
non-forward saturation model of Ref. \cite{MPS} (hereafter MPS model), which captures the main features of the dependence on energy,  virtual photon virtuality and momentum transfer $t$.  In the MPS model, the elementary elastic amplitude for dipole interaction is given by,
\begin{eqnarray}
\label{sigdipt}
\mathcal{A}_{q\bar q}(x,r,\Delta)= 2\pi R_p^2\,e^{-B|t|} {\cal{N}} \left(rQ_{\mathrm{sat}}(x,|t|),x\right),
\end{eqnarray}
with the asymptotic behaviors $Q_{\mathrm{sat}}^2(x,\Delta)\sim
\max(Q_0^2,\Delta^2)\,\exp[-\lambda \ln(x)]$. Specifically, the $t$ dependence of the saturation scale is parametrised as
\begin{eqnarray}
\label{qsatt}
Q_{\mathrm{sat}}^2\,(x,|t|)=Q_0^2(1+c|t|)\:\left(\frac{1}{x}\right)^{\lambda}\,, \end{eqnarray}
in order to interpolate smoothly between the small and intermediate transfer
regions. For the parameter $B$ we use the value $B=3.754$ GeV$^{-2}$ \cite{MPS}. Finally, the scaling function ${\cal{N}}$ is obtained from the forward saturation model \cite{IIM}.

Here, in order to take into account the skewedness correction, in the limit that $x^\prime \ll x \ll 1$, the elastic differential cross section should be multiplied by a factor $R_g^2$, given by \cite{Shuvaev:1999ce}
\begin{eqnarray}
\label{eq:Rg}
  R_g(\lambda_e) = \frac{2^{2\lambda_e+3}}{\sqrt{\pi}}\frac{\Gamma(\lambda_e+5/2)}{\Gamma(\lambda_e+4)}, \nonumber \\
  \quad\text{with} \quad \lambda_e \equiv \frac{\partial\ln\left[\mathcal{A}(x,\,Q^2,\,\Delta)\right]}{\partial\ln(1/x)},
\end{eqnarray}
which gives an important contribution mostly at large virtualities. In addition, we will take into account the correction for real part of the amplitude, using dispersion relations $Re {\cal A}/Im {\cal A}=\mathrm{tan}\,(\pi \lambda_e/2)$. In the MPS model, the skewedness correction is absorbed in the model parameters and only real part of amplitude will be considered.

\begin{figure}[h]
\includegraphics[scale=0.4]{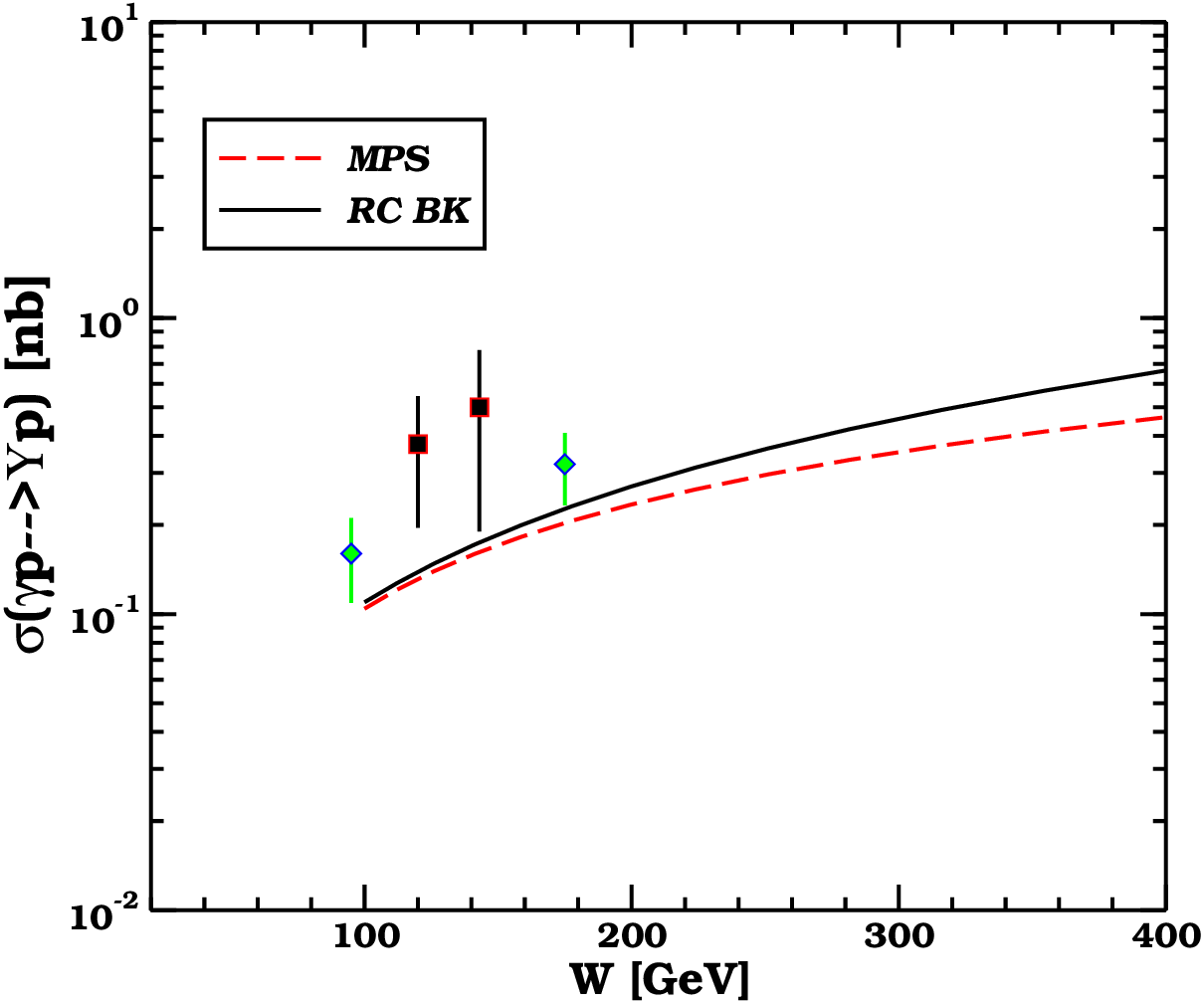}
\caption{Energy dependence of the $\gamma p$ cross section for $\Upsilon$ photoproduction. Data from ZEUS and H1 collaborations \cite{ZEUS_ups,H1_ups}.}
\label{fig:3}
\end{figure}

% (i.e., meson mass in photoproduction and photon virtuality in electroproduction)

Let us start to compare the RC BK predictions to the available HERA data for exclusive vector meson ($\rho$, $J/\Psi$ and $\Upsilon$) photo and electroproduction. In Fig. \ref{fig:1} we present the predictions of the RC BK and MPS models  for the diffractive $\rho^0$ vector meson production  and compare it with the current experimental data from ZEUS (left panel)  and H1 (right panel) Collaborations \cite{H1_rho,ZEUS_rho}.  These measurements are  interesting as they cover momenta scale  that are in the transition region between perturbative and nonperturbative physics, where saturation effects is expected to play an very important role. As the numerical RC BK solution there exists only for forward dipole-target amplitude we need an approximation to compute the non-forward amplitude. Here, we assume the usual exponential ansatz for the $t$-dependence which implies that the total cross-section is given by
\begin{eqnarray}
\sigma_{tot}(\gamma^*p\rightarrow Vp) = \frac{1}{B_V}\,\left. \left. \left[\frac{d\sigma_T}{dt}\right|_{t=0} + \frac{d\sigma_L}{dt}\right|_{t=0}\right]\,.
\end{eqnarray}
Notice that values of the slope parameter $B_V$ are not very accurately measured. We use the parametrisation
\begin{eqnarray}
B_V\,(Q^2) = 0.60\,\left[ \frac{14}{(Q^2+M_V^2)^{0.26}}+1  \right]
\end{eqnarray}
obtained from a fit to experimental data referred in Ref. \cite{code}. The uncertainty in this approximation can be larger than 20--30 $\%$ depending on the $Q^2$ value. It is verified that the effective power $\lambda_e$ is similar for both RC BK (solid line curves) and MPS (long dashed curves) predictions, with the deviation starting only at the higher $Q^2$ values where the predictions differ by a factor 1.5.  This can be a result of the similar small-$x$ behaviour for both models, where the effective power ranges from the soft Pomeron intercept $\lambda_e (Q^2=0)\approx \alpha_{\pom}(0)=1.08$ up to a hard QCD intercept $\lambda_e(Q^2) \simeq cN_c\alpha_s/\pi \approx 0.3$ for large $Q^2$. The data description is fairly good, with the main theoretical uncertainty associated to the choice of the light cone wavefunction (about a 15 $\%$ error).  It was verified that the contribution of real part of amplitude and skewedness are very small for $\rho$ production.

In Fig. \ref{fig:2} we present the  predictions of the RC BK model for the diffractive $J/\Psi$ production and compare with the ZEUS (left panel) and H1  (right panel) data \cite{H1_rho,ZEUS_rho}.  It is verified that the effective power $\lambda_e$ is similar for both RC BK and MPS only in the photoproduction case. The situation changes when the photon virtuality increases. The effective power for RC BK (solid line curves) is enhanced in $Q^2$ in comparison with the non-forward saturation model (long dashed curves).  The data description is reasonable since it is a parameter-free calculation and the uncertainties are similar as for $\rho$ production. For   $J/\Psi$ production, the contribution of real part of amplitude increase by 10 \%  the overall normalization, while the skewedness have a 20 \% effect. In the MPS model, as discussed before, the off-forward effects are absorbed in the parameters of model. The RC BK and MPS predictions  differ by a factor 1.4 for large energies. For sake of completeness, in Fig. \ref{fig:3} the results for $\Upsilon$ photoproduction is presented. The RC BK and MPS predictions are similar in the HERA energy range and differ by a factor 1.5 for large energies. It is known so far that the dipole approach underestimates the experimental data for $\Upsilon$. However, the deviation concerns only to overall normalization, whereas the energy dependence is fairly described. The referred enhancement in the effective power $\lambda_e$ is already evident in $\Upsilon$ photoproduction as the meson mass, $m_V = 9.46$ GeV, is a scale hard enough for deviations to be present. Skewedness is huge in the $\Upsilon$ case, giving a factor $R_g^2\approx 1.3$ in photoproduction. For this reason, we have included this effect in both models. However, this is not enough to bring the theoretical results closer to experimental measurements.

\begin{figure}[h]
\includegraphics[scale=0.4]{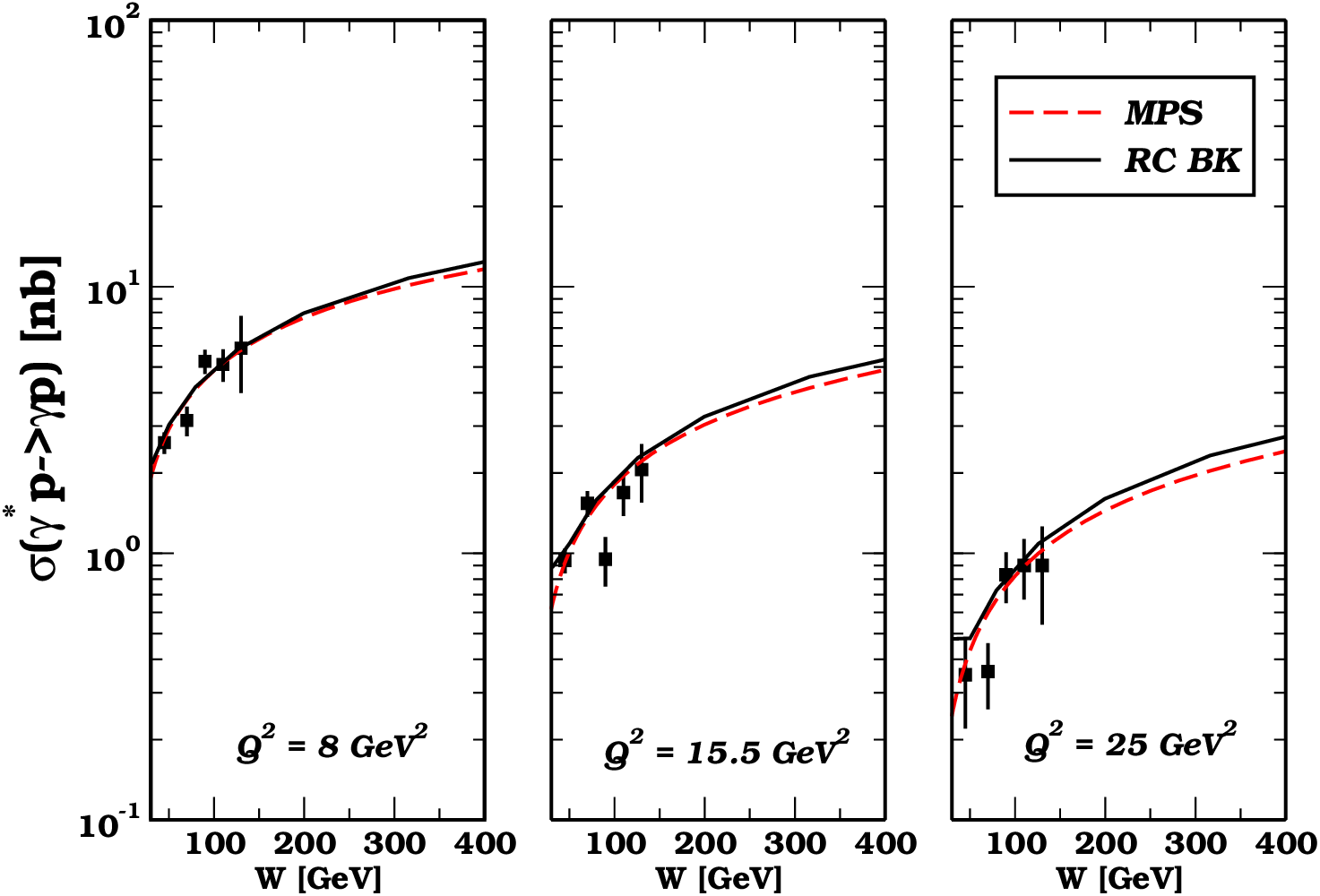}
\caption{Energy dependence of the DVCS cross section for different photon virtualities. Data from   H1 collaboration \cite{H1_dvcs}.}
\label{fig:4}
\end{figure}

Finally, we analyse the DVCS cross section and compare it to the recent H1 data \cite{H1_dvcs}. The cross sections are presented  as a function of $W$, for different values of $Q^2$, in Fig. \ref{fig:4}. Here, the approximations concerning the final state particle are not present and the cross section suffers of less uncertainties. For the slope value, we take the experimental parametrization \cite{H1_dvcs}, $B\,(Q^2)=a[1-b\log(Q^2/Q_0^2)]$, with $a=6.98 \pm 0.54 $ GeV$^2$, $b=0.12 \pm 0.03$ and $Q_0^2 = 2$ GeV$^2$. The situation for DVCS is similar as for vector meson photoproduction, where the effective power $\lambda_e$ is similar for both RC BK and MPS for small virtualities and starts to change as $Q^2$ grows. Skewedness is increasingly important for DVCS at high $Q^2$ and it was introduced for RC BK model. For the MPS model this effect is absorbed in the its parameters as noticed before. The RC BK and MPS predictions are similar for the HERA energy range, describing  the current data, and differ by a factor 1.2 for large energies.

%------------------------------------------------------------------------------

\section*{Acknowledgments}

It is a pleasure to thank the organizers and all participants for the
fruitful atmosphere in the school and workshop.

\end{document}